\documentstyle[amssymb,12pt,epsfig]{article}

\input{tcilatex}

\begin{document}

\baselineskip=18.6pt plus 0.2pt minus 0.1pt

\makeatletter
\@addtoreset{equation}{section} \renewcommand{\theequation}{\thesection.%
\arabic{equation}}

\begin{titlepage}

\title{
\hfill\parbox{4cm} {\normalsize Lab/UFR-HEP/02-02}\\ \vspace{1cm}
     {\bf   NC Geometry and Discrete Torsion\\
     Fractional Branes: I}
\author{
 El Hassan Saidi\thanks{{\tt H-saidi@fsr.ac.ma}} \ 
\\[3pt]
{ \small \it Sektion der Physik Theorie, Muenchen Universitat,
Muenchen }\\{ \small \it TheresienStr  37, D-80333, Muenchen,
Deutschland}\\
{ \small \it and}\\
{ \small \it Lab/UFR-High Energy Physics,  Department of Physics}\\
{ \small \it Faculty of Science, Av. Ibn Battouta, P.O Box 1014}\\
{ \small \it Rabat, Morocco}}} 
\maketitle
\begin{abstract}
Considering the complex $n$-dimension Calabi-Yau homogeneous
hyper-surfaces ${\cal H}_{n}$ and using algebraic geometry
methods, we develop the crossed product algebra  method introduced
 by Berenstein et Leigh in hep-th/0105229
and  build the non commutative (NC) geometries for
orbifolds ${\cal O}={\cal H}_{n}/{\bf Z}_{n+2}^{n}$ with a
discrete torsion matrix
$t_{ab}=exp[{\frac{i2\pi}{n+2}}{(\eta_{ab}-\eta_{ba})}]$, \
$\eta_{ab} \ \in SL(n,{\bf Z})$. We show that the NC manifolds
${\cal O}^{(nc)}$ are given by the algebra of functions on the
real $(2n+4)$ Fuzzy torus ${\cal T}^{2(n+2)}_{\beta_{ij}}$ with
deformation parameters
$\beta_{ij}=exp{\frac{i2\pi}{n+2}}{[(\eta^{-1}_{ab}-\eta^{-1}_{ba})}\
q_{i}^{a}\ q_{j}^{b}]$,  $q_{i}^{a}$'s being Calabi-Yau charges of
${\bf Z}_{n+2}^{n}$. We develop graph rules to represent ${\cal
O}^{(nc)}$ by quiver diagrams which become completely reducible at
singularities. Generic points in these NC geometries are be
represented by polygons with $(n+2)$ vertices linked by $(n+2)$
edges while singular ones are given by $(n+2)$ non connected
loops. We study  the various singular spaces of quintic  orbifolds  and  analyze the varieties of fractional $D$ branes at
singularities as well as the spectrum of massless fields. Explicit  solutions for the NC quintic ${\cal Q}^{(nc)}$ are derived
with details and general results for complex $n$ dimension
orbifolds with discrete torsion are presented.
\end{abstract}
\end{titlepage}
\thispagestyle{empty}
\newpage
\tableofcontents
\newpage

\section{Introduction}

\qquad Recently some interest has been given to build non commutative (NC)
extensions of Calabi-Yau orbifolds ${\cal O}$, with discrete torsion [1-10].
These NC manifolds, which could play some role in the twisted sector of
string compactification on orbifolds, go beyond the
standard non commutative ${\bf R}_{\theta }^{n}$ and NC$\ {\cal T}_{\theta
}^{k}$\ torii examples considered in brane physics [11-16]. They offer a
manner to resolve non geometric stringy singularities [17] and present a natural framework to study fractional $D$ branes at the singularities as well as
their massless field spectrum [18-23]. The basic idea of this construction
may be summarized as follows: the usual closed string compact target
subspace ${\cal O}$, which might be singular that is ${\cal O}\sim {\cal {M}}%
/G$ with $G\subset {\bf Z}_{m}^{k}$ having a set of fix points, is viewed as
the commuting central subalgebra ${\cal {Z}}\left( {\cal {A}}^{nc}\right) $
of a non commutative algebra ${\cal {A}}^{nc}$\ representing ${\cal O}^{nc}$%
, \ the NC complex manifold extending ${\cal {O}}$. \ The non commutative
algebra ${\cal {A}}^{nc}$ may be thought of as a fibration ${\bf B}\bowtie F$
\ with a base ${\bf B=}{\cal O}$ \ and as a fiber $F={\cal {A}}_{G}$ given by the
algebra of the group of discrete torsion $G$.

\qquad From the algebraic geometry point of view, \ the NC manifold \ ${\cal %
O}^{nc}$ \ is covered by a finite set of holomorphic matrix coordinate
patches ${\cal U}_{(\alpha )}=\{Z_{i}^{(\alpha )};1\leq i\leq n\;\alpha
=1,2,\ldots \}$ \ and holomorphic transition functions mapping \ ${\cal U}%
_{(\alpha )}$\ \ to \ ${\cal U}_{(\beta )}$; $\phi _{(\alpha ,\beta )}$: \ $%
{\cal U}_{(\alpha )}\ \ \rightarrow \ {\cal U}_{(\beta )}$. \ This is
equivalent to say that ${\cal {A}}^{nc}$ is covered by a collection of non
commutative local algebras ${\cal {A}}^{nc}{\cal {_{(\alpha )}}}$,
describing the analytic matrix coordinate patches ${\cal U}_{(\alpha )}$ of
the NC complex manifold, with analytic maps $\phi _{(\alpha ,\beta )}$ on
how to glue ${\cal {A}}_{(\alpha )}^{nc}$ and ${\cal {A}}^{nc}{\cal {%
_{(\beta )}}}$. The ${\cal {A}}^{nc}{\cal {_{(\alpha )}}}$ algebras have
centers ${\cal Z}_{\left( \alpha \right) }={\cal Z}\left( {\cal {A}}%
_{(\alpha )}^{nc}\right) $; \ when glued together give precisely the
commutative target space manifold ${\cal {O}}$. In this way a commutative
singularity of ${\cal {O}\simeq {Z}}\left( {\cal {A}}^{nc}\right) $ can be
made smooth in the non commutative space ${\cal {A}}^{nc}$ \cite{1,2}. This
idea was successfully used to build NC ALE spaces and some realizations of
Calabi-Yau threefolds such as the quintic threefolds ${\cal Q}$. In this
regards it was shown that the non commutative quintic ${\cal Q}^{nc}$
extending \ ${\cal Q}$,\ when expressed in the coordinate patch $%
Z_{5}=I_{id} $, \ is given by the following special algebra,
\begin{eqnarray}
Z_{1}Z_{2} &=&\alpha \ Z_{2}Z_{1},\qquad Z_{3}Z_{4}=\beta \gamma \
Z_{4}Z_{3},\ \ \ \ \ \ \ \ \ \ \ (a)  \nonumber \\
Z_{1}Z_{4} &=&\beta ^{-1}\ Z_{4}Z_{1},\qquad Z_{2}Z_{3}=\alpha \gamma \
Z_{3}Z_{2},\ \ \ \ \ \ \ \ (b)  \nonumber \\
Z_{2}Z_{4} &=&\gamma ^{-1}\ Z_{4}Z_{2},\qquad Z_{1}Z_{3}=\alpha ^{-1}\beta \
Z_{3}Z_{1},\ \ \ \ \ (c) \\
Z_{i}Z_{5} &=&\ Z_{5}Z_{i},\qquad i=1,2,3,4;  \nonumber
\end{eqnarray}
where $\alpha ,\beta $\ and\ $\gamma $\ are fifth roots of the unity, the
parameters of the ${\bf Z}_{5}^{3}$\ discrete group. The $Z_{i}$\ 's are the
generators of ${\cal {A}}^{nc}$. One of the main features of this non
commutative algebra is that its centre ${\cal Z(Q}^{nc})$ coincides exactly
with ${\cal Q}$, the commutative quintic threefolds. In \cite{2} , a special
solution for this algebra using $5\times 5$\ matrices has been obtained and
in \cite{24} a class of solutions for eqs(1.1) depending on the Calabi-Yau
charges of the quintic threefold has been worked out and some partial
results regarding higher dimensional Calabi-Yau hypersurfaces were given.
But open questions, such as the explicit dependence in the discrete torsion
of the orbifold group, the varieties of the fractional $D$ branes at
singularities on such kind of Calabi-Yau orbifolds and more generally fractional branes on NC toric manifolds, have not been answered
yet.

\qquad Moreover and as far eqs (1.1) are concerned, there are some first
remarks that one can make regarding the non commutative structure generated
from discrete torsion. Let us focus our attention on \ the first term of
the relations (1.1) namely
\begin{equation}
Z_{1}Z_{2}=\alpha \ Z_{2}Z_{1},
\end{equation}

 which, for the case of a complex dimension $n$ Calabi-Yau hypersurface with a ${\bf Z}_{n+2}$
discrete group, is supplemented equally by the conditions
\begin{eqnarray}
\alpha^{n+2}&=&1,\nonumber \\
Z_{1}^{n+2}&\sim& I_{id},\\
Z_{2}^{n+2}&\sim& I_{id}.\nonumber 
\end{eqnarray}
If for a while one forgets about the full set of eqs(1.1) and considers just the above relations, one discovers that they are nothing but the defining eq of a two dimensional fuzzy torus ${\cal T}_{\alpha }^{2}$ \ with a NC parameter $\alpha $. But where this NC fuzzy two torus comes from? What NC Calabi-Yau manifolds have to do with NC fuzzy torii? \par
 In $\ 4d$ $\ {\cal N}=4$ \ $U(N)$ super Yang-Mills
gauge theory where eqs type \  $\phi _{1}\phi _{2}=\alpha \ \phi _{2}\phi _{1}$ \ appear as vacuum configurations of strings on orbifolds preserving some
amount of supersymmetric charges, one suspects at first sight that  the NC  ${\cal T}%
_{\alpha }^{2}$\ might have origin in the toral subalgebra of the $U(N)$ gauge
group.\ But a  careful inspection shows  however that this is not true; after all NC calabi-Yau manifolds are mathematical objects and in general have nothing to do with super Yang-Mills theory. As we will show throughout this paper, the origin of  ${\cal T}%
_{\alpha }^{2}$\ is due to a nice property of $Z_{n+2}$ orbifolds; one of the two
NC cycles of ${\cal T}_{\alpha }^{2}$\ is related to the  ${\bf Z}_{n+2}$
discrete orbifold group while the other NC cycle is due an extra discrete hidden symmetry of the Calabi-Yau hypersurface orbifold. To avoid confusion with the original group, we  will  refer in this paper to the hidden invariance to as ${\bf Z}_{n+2}^{aut}$. As such the effective orbifold group should be thought as
${\bf Z}_{n+2}\ast {\bf Z}_{n+2}^{aut}$ rather than ${\bf Z}_{n+2}$ .
The $\ast$ product should not be confused with the star product of the Moyal plane one currently used in NC QFT a la Seiberg-Witten; the $\ast$ symbol is introduced here just to underline that the two ${\bf Z}_{n+2}$ and $%
{\bf Z}_{n+2}^{aut}$ groups do not commute.\ In fact, we will show that to
each ${\bf Z}_{n+2}$\ orbifold subsymmetry of the Calabi-Yau\ hypersurface
orbifold ${\cal H}_n / {\bf Z}_{n+2}^{n}$\ there exist indeed a ${\bf Z}%
_{n+2}^{aut}$ partner so that the full symmetry is $({\bf Z}_{n+2}\ast {\bf Z%
}_{n+2}^{aut})^{n}$. To fix the ideas, if we denote by $\left\{ \omega
_{i}\right\} $, the $\left( n+2\right) $ commuting group\ elements of \ $%
{\bf Z}_{n+2}$\, that is $\ \left\{ \omega _{i},\quad \omega _{i}\omega
_{j}=\omega _{j}\omega _{i};\quad \omega _{i}^{n+2}=I_{id}\right\} =\left\{
\omega _{i}=\omega ^{i};\quad \omega ^{n+2}=I_{id}\right\} $, and $\left\{
P_{\omega ^{i}}\right\} $ the elements of its representations ${\cal D}%
\left( {\bf Z}_{n+2}\right) $, $\ $then the$\ $\ ${\bf Z}_{n+2}^{aut}$
symmetry can be thought as a partner of ${\bf Z}_{n+2}$\ acting by rotating
the $\omega _{i}$ elements amongst themselves. In other words ${\bf Z}%
_{n+2}^{aut}$\ \ is just an automorphism group of ${\bf Z}_{n+2}$ operating
by adjoint actions. One of the remarkable features of ${\bf Z}_{n+2}^{aut}$\
is that it can be represented on the same space as that of ${\cal D}\left(
{\bf Z}_{n+2}\right) $\ and moreover does not commute with ${\bf Z}_{n+2}$\
; i.e ${\bf Z}_{n+2}\ast {\bf Z}_{n+2}^{aut}$ $\ \neq $ $\ {\bf Z}%
_{n+2}^{aut}\ast {\bf Z}_{n+2}$. Furthermore regarding usually the algebra
eqs(1.1) and too particularly the typical relation $Z_{1}Z_{2}=\alpha \
Z_{2}Z_{1}$ with eqs(1.3), one may ask why not NC geometries beyond the constraint eqs $%
\alpha ^{n+2}=1$, $Z_{1}^{n+2}\sim I_{id}$ and $Z_{2}^{n+2}\sim I_{id}$.
After all, the Fuzzy torus appearing in here is just a special realization of
the NC torus namely the rational representation; but there exists also an
irrational one, involving Powers Riffel projectors [25-27]. We suspect that such NC geometries exist and are associated with manifolds having a continuous NC  $U\left( 1\right) \ast
U_{aut}\left( 1\right) $ symmetry  which we believe to be  the kind of invariance that should appear in building NC toric manifolds. In this first paper, we will address the
rational case only, that is the situation where the $\alpha$ parameter is
such that $\alpha ^{n+2}=1$, by studying non commutative extensions of
orbifolds of complex dimension $n$ Calabi-Yau hypersurfaces with $\widehat{G}
$ symmetries contained in $\left( {\bf Z}_{n+2}\ast {\bf Z}%
_{n+2}^{aut}\right) ^{\otimes n}$. To do so we will need to go beyond the
constrained method initiated in [24] and use the algebraic geometry method
introduced in [17] which we will first develop it for the case of compact
Calabi-Yau manifolds; then we show that this method is a general one that
contain the constrained one. After that we  apply the new method to build the NC extensions
for the large class of Calabi-Yau hypersurfaces with explicit computations
for the example of the quintic. Next, using fibration ideas, we develop a way to get the singular spaces and  explore different kind of orbifolds
and study the various fractional $D$ branes at their singularities as well
as their massless field spectrum. In this paper, we shall also give explicit
realizations of discrete torsion using the automorphism symmetries of the
orbifold group and works out different classes of Morita equivalent
solutions for the NC geometry with torsion.

\qquad The organization of this paper is as follows: In section $2$, we give
the general setting for non commutative geometry from discrete
torsion by using algebraic geometry techniques. We first develop the
constrained method used in [24] and give the extension to higher dimensional
Calabi-Yau hypersurface orbifolds. Then we discuss an other realization
using results on $4d$ supersymmetric gauge theories where NC geometry
appears as the moduli space of vacuum configurations for superpotentials of
type \ \ $W\left( \Phi \right) =g\varepsilon ^{\lbrack IJK]}$ $\Phi _{I}\Phi
_{J}\Phi _{K}$. In section $3$, we give some general features of the quintic
symmetries and study its non commutative extensions using the constrained
method. Here we also complete partial results obtained in [24]. The analysis developed here applies as well to all Calabi-Yau hypersurfaces. In section $%
4 $, we develop non commutative geometry versus discrete torsion and work
out the full class of torsion dependent solutions. In section $5$, we
develop the crossed product (CP) algebra method introduced in [17] and use this development to derive general solutions for the NC Calabi-Yau orbifolds. This
analysis recovers the results of [17] as special cases. In section $6$, we
study the symmetries of the moduli space solutions we have obtained and in
section $7$, we discuss the varieties of fractional $D$ branes and give the
full spectrum of massless fields on the $D $ branes. In section $8$, we
collect some general results on NC complex $d$-dimensional Calabi-Yau
orbifolds with a discrete torsion matrix $t_{ab}=exp[{\frac{i2\pi }{d+2}}{%
(\eta _{ab}-\eta _{ba})}]$, \ $\eta _{ab}\ \in SL(d,{\bf Z})$ and study the
classes of varieties of fractional D branes at singularities. The NC
manifolds are given by the algebra of functions on the Fuzzy torus ${\cal T}%
_{\beta _{ab}}^{2d}$, where $\beta _{ab}=exp{\frac{i2\pi }{d+2}}{[(\eta
_{cd}^{-1}-\eta _{dc}^{-1})}\ q_{a}^{c}\ q_{b}^{d}]$ . Finally we give a
conclusion.

\section{General Setting}

\qquad Given a commutative Calabi-Yau manifold ${\cal M}$, in particular
that class of orbifolds ${\cal O}$ with discrete torsion group \ $G$, \ $%
G\subseteq {\bf Z}_{n+2}^{n}$, \ described by complex dimension $n$
commutative holomorphic $(n+2)$ homogeneous hypersurface ${\cal H}_{n+2}=%
{\cal P}_{n+2}(z_{1},\ldots ,z_{n+2})=\lambda^{-n-2}{\cal P}_{n+2}(\lambda z_{1},\ldots ,\lambda z_{n+2})$, embedded in the complex projective
space ${\bf CP}^{n+1}$, \ one can build its non commutative geometry
extension ${\cal O}^{nc}$ by using discrete torsion cocycles. In this section, we want to show that in general there are two
ways one may follows to achieve this goal: $(1)$ A constrained approach
using purely algebraic computations and $(2)$ the CP algebra method based on
techniques of algebraic geometry and representation theory of $G$. For
pedagogical reasons, we shall refer here below to the first method as the $Z$
realization because of its direct connection with holomorphic analysis of
complex manifolds and to the second one as the $\Phi $ representation due to
its direct physical origin; The CP algebra describes vacuum configurations
of some supersymmetric field theories on orbifolds, . We will show that
these two realizations are in fact two special basis of the algebra of
functionals on fuzzy torii \ ${\cal T}_{{\bf \beta }_{ij}}^{2\left(
n+2\right) }$ \ and are linked to each other by a Morita morphism. In what
follows, we develop the general setting of the methods with special focus on
the basic features of each one of the two realizations; more developments
and explicit solutions of both of them will be given in the next sections.

\subsection{$Z$ Representation}

\qquad If we denote by ${\cal R}_{n+2},$ the ring of holomorphic and degree $%
(n+2)$ homogeneous polynomials on ${\bf CP}^{n+1}$, namely ${\cal P}{%
(z_{1},\dots ,z_{n+2})}=z_{1}^{n+2}+...+z_{n+2}^{n+2}+a_{0}%
\prod_{i=1}^{n+2}z_{i}=0,$ and by $\ {\cal R}_{n+2}^{\left[ \alpha \right] }
\equiv \ {\cal R}_{n+2}/G_{\left[ \alpha \right] }$, \ the set of equivalent
classes of homogeneous polynomials that are invariant under the discrete
group $G_{\left[ \alpha \right] }\subseteq {\bf Z}_{n+2}^{n}$, then one can
build the  non commutative extension of $\ {\cal R}_{n+2}^{\left[ \alpha \right] }$   by using $G_{\left[ \alpha \right] }$
discrete torsion cocycles. There are essentially two key steps in using this
constrained construction: (1) A correspondence rule which associate to each
complex variable $z$, a chiral matrix operator $Z$ valued in some $G_{\left[ \alpha \right] }$  representation  and (2) a set of
constraint eqs ensuring that the original commutative manifold sit in the
centre of the NC space. In what follows, we summarize the main lines of this
construction.

\subsubsection{Correspondence Rule: Commutative $\longrightarrow $ Non
Commutative}

\qquad Starting from the polynomial ${\cal P}{(z_{1},\dots ,z_{n+2})}$, we
associate to each commutative complex \ holomorphic variable $z_{i}$, a
(matrix) chiral operator $Z_{i}$ valued in some representation ${\cal D}%
\left( G\right) $ of the orbifold group $G$. Formally, if \ $\left\{
T_{\{a\}};\qquad a=1,...\right\} $ \ denotes the set of the elements (
generators) of the ${\cal D}\left( G\right) $ representation satisfying the $%
G$ group law,
\begin{eqnarray}
T_{\{a\}}\ T_{\{b\}}\ T_{\{a\}}^{-1}\ &=&\ t_{\{a,b\}}\ T_{\{b\}},  \nonumber
\\
T_{\{a\}}^{n+2} &=&I_{{\cal D}\left( G\right) }
\end{eqnarray}
with a given non trivial torsion matrix $t_{ab},$ and acting on the $z_{i}$
variables with eigencharges $\lambda _{i\{a\}}$,
\begin{eqnarray}
T_{\{a\}}\ z_{i}\ T_{\{a\}}^{-1}\ &=&\ \lambda _{i\{a\}}\ z_{i},  \nonumber
\\
T_{\{a\}}^{n+2}\ z_{i}\ T_{\{a\}}^{-\left( n+2\right) }\ &=&\ \ z_{i}
\end{eqnarray}
then the correspondence rule Commutative-Non commutative, reads as;
\begin{eqnarray}
z_{i}\ \ &\rightarrow &\ \ \ Z_{i}=\prod_{a}Z_{ia_{i}}  \nonumber \\
Z_{ia_{i}} &=&x_{i}\otimes \ T_{\{a_{i}\}}\equiv x_{i}\ T_{\{a_{i}\}}.
\end{eqnarray}
Here the $x_{i}$\ parameters are ${\bf C}$-numbers neutral under the $T_{\{a\}}$
\ actions and the $\prod_{a}$  product is carried on the $T_{\{a_{i}\}}$  .\ From this correspondence, one sees that the $Z_{i}$ chiral
operators inherit features of the ${\cal D}\left( G\right) $ group
representations. In particular, we have for the commutation relations of the
$Z_{ia_{i}}$\ , the same cocycles as those appearing in the commutations of
the $T_{\{a_{i}\}}$'s \ as shown here below:
\begin{eqnarray}
Z_{ia_{i}}\ Z_{jb_{j}} &=&\ t_{\{a_{i},b_{j}\}}\ Z_{jb_{j}}Z_{ia_{i}},
\nonumber \\
T_{\{a_{i}\}}T_{\{b_{j}\}} &=&\ t_{\{a_{i},a_{j}\}}\ T_{\{bj\}}T_{\{a_{i}\}},
\\
Z_{i}\ Z_{j} &=&\ t_{ij}\ Z_{j}Z_{i},  \nonumber
\end{eqnarray}
with $\ t_{ij}=\prod_{a,b}t_{\{a_{i},b_{j}\}}$ \ These relations show
clearly that non commutative geometry, as built above, is induced by
discrete torsion and has much to do with fuzzy realization of non
commutative torii since the above relations look like the familiar $%
UV=\lambda VU$ equation of the two dimensional \ NC torus $[11,27]$. We will
turn to this special issue in the development; for the moment let us  comment
briefly the above relations. Eqs(2.1) are general relations which may be
exploited in different ways. Just for the simple example where $G={\bf Z}%
_{n+2}$ \ where we have one apparent generator $T_{1}=P_{\omega }\equiv P$,
one can build various torsion matrices ${\bf t}=(t_{ab})$. Besides the usual $t_{11}=1$  term, one can build several non trivial ones. Taking into account the ${\bf Z}%
_{n+2}^{aut}$\ automorphism group generated by $Q$ and setting $%
T_{2}=Q_{\omega }=Q$, the ${\bf t}$ matrix reads as
\begin{equation}
t_{ab}=\left(
\begin{array}{cc}
1 & \omega ^{-1} \\
\omega & 1
\end{array}
\right) .
\end{equation}

An other example involving the $\widehat{{\bf Z}_{n+2}}\equiv {\bf Z}%
_{n+2}\ast {\bf Z}_{n+2}^{aut}$ \ \ group, is to take the $T_{a}$ operators as
given by elements of ${\bf Z}_{n+2}$ and ${\bf Z}_{n+2}^{aut}$. Generic
examples correspond to take $T_{a}$\ as given by $r$ elements of ${\bf Z}%
_{n+2}$\ and $m$ elements of ${\bf Z}_{n+2}^{aut}$; i.e $\left\{
T_{a}\right\} =\left\{ P_{\omega _{1}},...,P_{\omega _{r}};Q_{\omega
_{1}},...,Q_{\omega _{m}}\right\} $. In this case torsion is carried by a $%
\left( r+m\right) \times \left( r+m\right) $ matrix given by,
\begin{equation}
t_{ab}=\left(
\begin{array}{cc}
\left( I_{id}\right) _{r\times r} & \left( A\right) _{r\times m} \\
\left( B\right) _{m\times r} & \left( I_{id}\right) _{m\times m}
\end{array}
\right) ,
\end{equation}
where the rectangular $A$ and $B$ matrices read as,
\begin{eqnarray}
A &=&\left(
\begin{array}{ccccc}
\omega ^{-1} & \omega ^{-1} & ... & \omega ^{-1} & \omega ^{-1} \\
\omega ^{-2} & \omega ^{-2} & ... & \omega ^{-2} & \omega ^{-2} \\
\begin{array}{c}
. \\
.
\end{array}
&
\begin{array}{c}
. \\
.
\end{array}
&
\begin{array}{c}
... \\
...
\end{array}
&
\begin{array}{c}
. \\
.
\end{array}
&
\begin{array}{c}
. \\
.
\end{array}
\\
\omega ^{1-m} & \omega ^{1-m} & ... & \omega ^{1-m} & \omega ^{1-m} \\
\omega ^{-m} & \omega ^{-m} & ... & \omega ^{-m} & \omega ^{-m}
\end{array}
\right) _{r\times m},  \nonumber \\
&&  \nonumber \\
&& \\
B &=&\left(
\begin{array}{ccccc}
\omega & \omega ^{2} & ... & \omega ^{m-1} & \omega ^{m} \\
\omega & \omega ^{2} & ... & \omega ^{m-1} & \omega ^{m} \\
\begin{array}{c}
. \\
.
\end{array}
&
\begin{array}{c}
. \\
.
\end{array}
&
\begin{array}{c}
... \\
...
\end{array}
&
\begin{array}{c}
. \\
.
\end{array}
&
\begin{array}{c}
. \\
.
\end{array}
\\
\omega & \omega ^{2} & ... & \omega ^{m-1} & \omega ^{m} \\
\omega & \omega ^{2} & ... & \omega ^{m-1} & \omega ^{m}
\end{array}
\right) _{m\times r}.  \nonumber
\end{eqnarray}

It is this kind of matrix torsion which has been used in [24] to build a
class of solutions for the NC quintic. Later on we will use an other kind of
torsion matrix by focusing our attention on the basic $P$\ and $Q$
generators of the $\widehat{{\bf Z}_{n+2}^{k}}\equiv \left( {\bf Z}%
_{n+2}\ast {\bf Z}_{n+2}^{aut}\right) ^{\otimes k}$\ group. To avoid
confusion, we will refer to the basic $k$ generators $P_{\omega }$ of ${\bf Z%
}_{n+2}^{k}$ (respectively $Q_{\omega }$ of $\left( {\bf Z}%
_{n+2}^{aut}\right) ^{k}$) as $E_{a}$, $a=1,...,k$ ( respectively $J_{a}$).

Concerning eqs(2.3), it is interesting to note that in addition to the ${\bf %
Z}_{n+2}^{n}$\ symmetries eqs(2.2), the $z_{i}$\ complex variables carry
equally a $U(1)$ charge of the complex space describing rotations in ${\bf R}%
^{2}\simeq {\bf C}$\ and acting as $z\longrightarrow z\exp i\chi $, with $%
\chi $\ a continuous parameter. The splitting eq(2.3) consists to separate
the ${\bf Z}_{n+2}^{n}$ and the $U(1)$\ charges. The $x_{i}$ variables carry
now the $U(1)$\ charge only while the ${\bf Z}_{n+2}^{n}$\ charges are
carried by the $T_{a}$'s; that is,
\begin{eqnarray}
T_{\{a\}}\ x_{i}\ T_{\{a\}}^{-1}\ &=&\ x_{i},  \nonumber \\
T_{\{a\}}\ T_{b}\ T_{\{a\}}^{-1}\ &=&\ \lambda _{i\{b\}}\ T_{b}.
\end{eqnarray}

In practice, and as we will use in section 6, we shall often isolate only
the $G_{\left[ \alpha \right]}={\bf Z}_{n+2}^{ \alpha}$ charges of interest;
that is the charges of orbifold subgroup $G_{\left[ \alpha \right] }$, and
so the $x_{i}$ complex variables still carry the charges of $G/G_{\left[
\alpha \right] }$. In other words $T_{\{a\}}\ x_{i}\ T_{\{a\}}^{-1}\ =\
x_{i} $ for \ $T_{\{a\}}\in G_{\left[ \alpha \right] }$\ and $T_{\{a\}}\
x_{i}\ T_{\{a\}}^{-1}\ =$ $\lambda _{i\{a\}}\ x_{i}$ for \ $T_{\{a\}}\in
G/G_{\left[ \alpha \right] }$.

\subsubsection{Constraint Eqs: \ Commutative Geometry is in the centre of
the NC algebra.}

\qquad Eq(2.4) tells us that the $Z_{ia_{i}}$ chiral matrix operators may be
viewed as the NC generators of a functional algebra ${\cal A}_{nc}\left[ Z%
\right] $,
\begin{equation}
{\cal A}_{nc}\left[ Z\right] \equiv {\cal A}\left[ Z_{1a_{1}},\ldots
Z_{\left( n+2\right) a_{n+2}}\right],
\end{equation}
living on a commutative orbifold parameterized by the $x_{i}$ complex
variables, but valued in the algebra ${\cal A}_{{\cal D}\left( G\right) }$
of the group representation ${\cal D}\left( G\right) $. In other words $%
{\cal A}_{nc}\left[ Z\right] $\ may be thought of as a fiber bundle \ ${\cal %
B}\boxtimes {\cal F}$ \ whose base \ ${\cal B}$ \ is \ ${\cal R}_{n+2}/G_{%
\left[ \alpha \right] }$ \ and its fiber \ ${\cal F}$ \ is just \ ${\cal A}_{%
{\cal D}\left( G\right) }$.
\begin{equation}
{\cal A}_{nc}\left[ Z\right] \sim \left( {\cal R}_{n+2}/G_{\left[ \alpha %
\right] }\right) \boxtimes {\cal A}_{{\cal D}\left( G\right) }.
\end{equation}

The commutative orbifold coincide therefore with the centre ${\cal Z}\left(
{\cal A}_{nc}\left[ Z\right] \right) $ of the NC algebra ${\cal A}_{nc}\left[
Z\right] .$%
\begin{equation}
{\cal Z}\left( {\cal A}_{nc}\left[ Z\right] \right) \equiv {\cal R}_{n+2}/G_{%
\left[ \alpha \right] }
\end{equation}

From this property, it follows the constraint relations
\begin{eqnarray}
\left[ Z_{ia_{i}},Z_{ja_{j}}^{n+2}\right] &=&0,  \nonumber \\
\left[ Z_{ia_{i}},\prod_{j=1}^{n+2}\ Z_{ja_{j}}\right] &=&0.
\end{eqnarray}

The first constraint eq reflects just the group property carried by the
second relations of eqs(2.1) and (2.2) while the second constraint has much
to  do with the Calabi-Yau condition and turns out to play a central role
in working out the solutions of the $Z_{ia_{i}}$\ generators.

Using the correspondence eqs(2.3) and by help of the Schur lemma of group
representation theory, one sees that the set $\{T_{\{a_{i}\}}\}$ of the
group representation generators appearing in eq(2.3) and the cocycles \ $%
t_{\{a_{i},a_{j}\}}$\ of eq(2.4) are determined by finding those $%
T_{\{a_{i}\}}$ elements satisfying,
\begin{eqnarray}
T_{\{a_{i}\}}^{n+2} &=&I_{{\cal D}\left( G\right) },  \nonumber \\
\prod_{j=1}^{n+2}\ T_{a_{j}}\ &=&I_{{\cal D}\left( G\right) }.
\end{eqnarray}
\

\qquad From this brief presentation, one sees that one can get the non
commutative extension of ${\cal R}_{n+2}/G$ directly by solving the
constraint eqs (2.12-13) without need to introduce powerful tools using more
involved analysis on the fibration eqs(2.10). This is a strong point in
favor of the constrained method; especially that in the $Z$
realization, non commutative geometry is manifest as shown on eqs(2.4).
There are however weak points for this constrained method; features such as
the fractional $D$ branes at singularities, the explicit dependence of the
deformation parameters into the discrete torsion matrix ${\bf t}_{ab}$ and
hidden quantum symmetries are not completely manifest. Some special issues
of the $Z$ representation has been first considered in $[2]$, and
re-examined in $[24]$. Here we will develop the general setting of the
method, work out general solutions exhibiting larger symmetries and give
further explicit details for the case of the non commutative quintic and
more generally for the ${\cal H}_{n}$ hypersurfaces.

\subsection{$\Phi $ \ Representation}

\qquad Building non commutative extensions of Calabi-Yau orbifolds ${\cal O}$
with discrete torsion can be achieved by using a more involved method; the
CP algebra considered recently in $[17]$. To get the idea behind this
approach and how it appears in supersymmetric field theory embedded in
string theory, let us start from the following special chiral superpotential
\begin{equation}
W\left( \Phi \right) =g\varepsilon ^{\lbrack IJK]}\text{ }\Phi _{I}\Phi
_{J}\Phi _{K}
\end{equation}
of four dimension Super Yang-Mills theories; with $g$ a coupling constant, $%
\varepsilon ^{\lbrack IJK]}$\ is the usual completely antisymmetric and $%
\Phi =\phi +\theta \psi +\theta \overline{\theta }F$, the $4d$ chiral
superfields with $\theta $ and $\overline{\theta }$ the well known $4d$
superspace variables [29].

This \ $W(\Phi )$\ superpotential \ has a manifest ${\bf Z}_{k}\otimes {\bf Z%
}_{k}$ symmetry, with discrete torsion, respectively generated by ${\bf e}$
and ${\bf f}$ operators and acting as,
\begin{eqnarray}
{\bf e}\ \Phi _{I}\ {\bf e}^{-1}\ &=&\ \Phi {_{I}}\ exp\ ({\frac{i2\pi
q_{I}^{e}}{k}})\ ,  \nonumber \\
{\bf f}\ \Phi _{I}\ {\bf f}^{-1}\ &=&\ \Phi _{I}\ exp\ ({\frac{i2\pi
q_{I}^{f}}{k}}),  \nonumber \\
{\bf f}\ {\bf e}\ {\bf f}^{-1}\ &=&\ t_{fe}\ {\bf e},\quad {\bf e}%
^{k}=I_{id},\quad {\bf f}^{k}=I_{id},
\end{eqnarray}

where the $q_{I}^{e}$ and $q_{I}^{f}$ charges are such that;

\begin{equation}
\sum_{I=1}^{3}\ q_{I}^{e}=0,\qquad \sum_{I=1}^{3}\ q_{I}^{f}=\ 0.
\end{equation}

\qquad The form eq(2.14) of the superpotential is very special; it is
intimately connected with the Lie algebra invariant form $tr\left[ A,\left[
B,C\right] \right] $ and plays an important role in different supersymmetric
theories. First, it appears in $4d$ \ ${\cal N}=4$ \ supersymmetric field
theory with a $U(N)$ gauge group, containing, in addition to the fermions
and the $4d$ gauge field $A_{\mu }$, six scalars fields transforming in the
representation \ $\underline{{\bf 6}}{\bf \times }\left( \underline{{\bf N}}%
{\bf \times }\overline{{\bf N}}\right) $ of $\ SO(6)\times U(N)\simeq
SU(4)\times U(N)$ symmetry. In language of superfields, this corresponds to
interpret $\Phi _{I}$ as $\Phi _{i}=\sum_{\alpha =1}^{N^{2}}\Phi _{i\alpha }%
{\sl T}^{\alpha }$, $i=1,...,6$, the coupling parameter $g$ as the $4d$ \ $%
{\cal N}=4$ Yang-Mills coupling constant and finally $\varepsilon ^{\lbrack
IJK]}$ as $f^{\alpha \beta \gamma }$, \ the completely antisymmetric $SU(N)$
Lie algebra structure constants. Put differently, the superpotential of the $%
4d$ \ ${\cal N}=4$\ SYM expressed in terms of the $4d$ \ ${\cal N}=1$ \
superfields is $W\left( \Phi \right) \sim g_{SYM}^{N=4}\ Tr\left[ \Phi _{i},%
\left[ \Phi _{j},\Phi _{k}\right] \right] $.

\qquad Eq(2.14) may be also thought of as the \ superpotential of the pure $%
4d$ ${\cal N}=2\ \ SU(2)$ gauge model where the three $\Phi _{I}$ chiral
superfields are taken in the adjoint of $SU(2)$, namely $\Phi
_{I}=\sum_{i=1}^{N^{2}}\Phi _{i}$ $\sigma ^{i}$ , with $\sigma ^{\alpha }$\
the usual Pauli matrices, which together with the gauge multiplet form a $4d$
${\cal N}=2\ \ SU(2)$ super Yang-Mills representation. In this case, $g$ is
the $4d$ \ ${\cal N}=2$ Yang-Mills coupling constant $g_{SYM}^{N=2}$ and $%
\varepsilon ^{\lbrack IJK]}$ is the usual completely antisymmetric $SU(2)$
structure constants $\varepsilon ^{ijk}$. Recall in passing that both \ the $%
4d$ ${\cal N}=4\ \ $and ${\cal N}=2$ field models may be viewed as the low
energy effective field theory of $10d$ strings compactifications on some
large volume Calabi-Yau \ threefolds, such as the six torus $T^{6}$\ or
again the Calabi-Yau elliptic fibrations $K_{3}\bowtie T^{2}$ and \ ${\bf CP}%
^{1}\bowtie {\bf CP}^{1}\bowtie T^{2}$.

\qquad An other example where the superpotential $W\left( \Phi \right)
=g\varepsilon ^{\lbrack IJK]}$ $\Phi _{I}\Phi _{J}\Phi _{K}$ \ appears is in
the low energy field theory of the $10d$ string on $AdS_{5}\times S^{5}/{\bf %
Z}_{3}$ \ \ near the horizon, with a system of $N$ $\ $type IIB$\;D3$ branes
located at the orbifold points of $AdS_{5}\times {\ S^{5}}/ {\Gamma} \sim
{\bf R}^{1,3}\times {\bf R}^{6}/{\Gamma }$. The orbifold discrete group $%
\Gamma $ is in the $SO_{R}(6)\sim SU_{R}(4)$ $\ R$-symmetry and according to
whether $\Gamma =I_{id}$, or $\Gamma \subset SU(2)\subset SU_{R}(4)$ or
again $\Gamma \subset SU(3)\subset SU_{R}(4)$, one gets either $4d$ \ ${\cal %
N}=4$ or $\ 4d\ {\cal N}=2$ or again $4d$ \ ${\cal N}=1$ \ supersymmetric
theories respectively [30-31]. In the case where $\Gamma ={\bf Z}_{3}\subset
SU(3)$ for instance, the three sets of complex transverse bosonic
coordinates $\phi _{I}$ of \ the $N$ $D3$ branes on ${\bf C}^{3}/{\bf Z}_{3}$%
, are just the scalar components of chiral superfields $\Phi _{i}$
transforming in the bi-fundamentals of $SU\left( N\right) ^{3}$; i.e $\ \Phi
_{1}\sim \left( N,\overline{N},1\right) $, $\Phi _{2}\sim \left( 1,N,%
\overline{N}\right) $ and $\ \Phi _{3}\sim \left( \overline{N},1,N\right) $.
For details see [32].

\qquad The next step in this field analysis is to look for supersymmetric
vacuum configurations of the supersymmetric scalar potential $V\left( \phi
\right) $ induced by eliminating the supersymmetric $F$ auxiliary fields.
Upon expressing these auxiliary fields $F$ in terms of the $\phi $, the
vanishing condition of the scalar potential is $V\left( \phi \right)
=\sum_{I=1}^{3}\left| \frac{\partial W}{\partial \phi _{L}}\right| ^{2}=0$
reads in general as;
\begin{equation}
\sum_{I,J=1}^{3}\ \left[ \phi _{J}\phi _{K}-\phi _{K}\phi _{J}\right] ^{2}=0%
\text{.}
\end{equation}

For complex three dimension orbifolds ${\cal M}/{\bf Z}_{3}\otimes {\bf Z}%
_{3}$ with discrete torsion, the above condition, which can also be
expressed as $\phi _{J}\phi _{K}-\phi _{K}\phi _{J}=0$, have to be
supplemented with the symmetries eqs(2.15). Putting all these things
together, the supersymmetric vacuum configurations, for string
compactifications on ${\bf Z}_{3}\otimes {\bf Z}_{3}$ orbifolds with
discrete torsions, reads then as:
\begin{eqnarray}
\phi _{J}\phi _{K} &=&\phi _{K}\phi _{J}  \nonumber \\
{\bf e}\ \phi _{j}\ {\bf e}^{-1}\ &=&\ \phi {_{j}}\ exp\ ({\frac{i2\pi
q_{j}^{e}}{3}})\ ,  \nonumber \\
{\bf f}\ \phi _{j}\ {\bf f}^{-1}\ &=&\ \phi _{j}\ exp\ ({\frac{i2\pi
q_{j}^{f}}{3})} \\
{\bf f}\ {\bf e}\ {\bf f}^{-1}\ &=&\ t_{fe}\ {\bf e}.  \nonumber
\end{eqnarray}

Setting $e_{a}=(e,f)$ \ and \ $\ \lambda _{ai}=(exp({\frac{i2\pi q_{i}^{e}}{3%
}});\ exp({\frac{i2\pi q_{i}^{f}}{3}}))$, \ we can rewrite all above
informations on supersymmetric vacuua on ${\bf Z}_{3}\otimes {\bf Z}_{3}$
orbifolds into the following condensed form

\begin{eqnarray}
\Phi _{i}.\Phi _{j}\ &=&\ \Phi _{i}.\Phi _{j},  \nonumber \\
e_{a}\Phi _{i}\ &=&\ \lambda _{ai}\ \Phi _{i}e_{a}, \\
e_{a}\ e_{b}\ &=&\ t_{ab}\ e_{b}\ e_{a},  \nonumber
\end{eqnarray}

where we have used capital Greek letters to represent the bosonic scalar
components of the chiral fields; they should not be confused with
superfields. Actually these relations generate a non commutative crossed
product algebra of ${\bf C}^{3}{\bf /Z}_{3}\otimes {\bf Z}_{3}$. This CP
algebra will be denoted as \ \ ${\cal O}\ \boxtimes \ {\cal A}_{G}$, with $%
{\cal O}={\bf C}^{3}{\bf /}G$\ and $G={\bf Z}_{3}^{2}$; it can be thought
of as a fibration of the algebra \ ${\cal A}_{G}$ \ of the group $G$ on $%
{\cal O}\ $. We will refer hereafter to eqs(2.19) as the $\Phi $\
representation of the non commutative orbifold, \ ${\cal O}^{(nc)}\equiv
{\cal O}\ \boxtimes \ {\cal A}_{G}$,\ \ and denote it as \ ${\cal A}[\Phi
_{1},\Phi _{2},\Phi _{3};{\bf e}_{1},{\bf e}_{2}]$ or ${\cal A}[\Phi ;{\bf e}%
]$\ for short. The latter, which is just the algebra of functionals on \ $%
{\cal O}$, is in fact a remarkable basis of \ \ ${\cal O}\ \boxtimes \ {\cal %
A}_{G}$; An other different, but equivalent, basis of ${\cal O}^{(nc)}$\ \
is that given by the following $Z$\ realization to which we will also refer
to as \ ${\cal A}[Z_{1},Z_{2},Z_{3};e,f]$ or ${\cal A}[Z;{\bf e}]$ for short,

\begin{eqnarray}
Z_{i}.Z_{j}\ &=&\ \beta _{ij}\ Z_{i}.Z_{j},  \nonumber \\
{\bf e}_{a}.Z_{i}\ &=&\ \ Z_{i}.{\bf e}_{a}, \\
{\bf e}_{a}\ {\bf e}_{b}\ &=&\ t_{ab}\ \ {\bf e}_{b}\ {\bf e}_{a}.  \nonumber
\end{eqnarray}

The $\beta _{ij}$'s are some parameters such as $\beta _{ij}^{3}=1$; their explicit relation whith torsion will be given later on.
In these equations, the chiral operators $Z_{i}$ are related to the $\Phi
_{i} $ ones by a  ${\bf e}_{a}$ dependent ( Morita) morphism  $\vartheta _{i}$;relating the two descriptions; i.e
\ $Z_{i}=\vartheta _{i}({e_{a}})\ \Phi _{i}$. More details on this change of
frames will be discussed later on; see section 7.

{\sl Extension}

\qquad Given a complex $n$-dimension orbifold ${\cal O}_{n}={\cal M}%
_{n}/G_{m}$ with $G_{m},$ a dimension $m$ discrete group, the $\Phi $ basis,
\ ${\cal A}\left[ \Phi _{1},\ldots ,\Phi _{n};F_{1},\ldots ,F_{m}\right]
\equiv {\cal A}\left[ \Phi ;G_{m}\right] $, \ of the non commutative CP
algebra \ ${\cal O}_{n}\ \boxtimes \ A_{G_{m}}$ \ is given by the following
generalization of eqs(2.19), namely
\begin{eqnarray}
\Phi _{i}.\Phi _{j}\ &=&\ \Phi _{i}.\Phi _{j},\ \ \ i,j=1,\ldots ,n,
\nonumber \\
F_{a}\Phi _{i}\ &=&\ \lambda _{ai}\ \Phi _{i}F_{a}, \\
F_{a}\ F_{b}\ &=&\ t_{ab}\ F_{b}\ F_{a},\ \ \ a,b=1,\ldots ,m.  \nonumber
\end{eqnarray}
By an appropriate change of basis $Z_{i}=\vartheta _{i}({e_{a}})\ \Phi _{i}$%
, one can also define the $Z$ representation ${\cal A}\left[ Z;G_{m}\right] $
of the non commutative CP space \ ${\cal O}_{n}\boxtimes A_{G_{m}}$ as the
algebra generated by the following relations;
\begin{eqnarray}
Z_{i}\ Z_{j}\ &=&\ \beta _{ij}\ \ Z_{i}\ Z_{j},\ \ \ i,j=1,\ldots ,n,
\nonumber \\
F_{a}\ Z_{i}\ &=&\ Z_{i}\ F_{a}, \\
F_{a}\ F_{b}\ &=&\ t_{ab}\ F_{b}\ F_{a},\ \ \ a,b=1,\ldots ,m,  \nonumber
\end{eqnarray}
where now $\beta _{ij}$ are non trivial cocycles. The difference between
eqs(2.21-22) is just an apparent one; both of the $\Phi $\ and $Z$ frames
are in fact Morita equivalent. The key point in the non similarities between
the two representations is that in the $\Phi $ frame the non commutative
orbifold ${\cal O}_{n}^{(nc)}$ is given by a non trivial fibration of the
algebra ${\cal A}_{G}$ on ${\cal Z}\left( {\cal A}\left[ \Phi ;G_{m}\right]
\right) $; while this fibration becomes trivial in the $Z$ realization. Both
$\Phi $ and $Z$ describe the NC geometry extension of the same orbifold
though non commutative geometry is non manifest on the $\Phi _{i}$'s. We
shall show later on that the $\Phi $ representation has however some
remarkable features whose main ones are:

(i) fractional $D$ branes phenomenon is manifest in the $\Phi $ realization;
it is connected with complete reducibility of the representation at
singularities. In other words, we will show that the non zero $\Phi _{i}$
matrix brane transverse coordinates decompose at the orbifold points as
follows
\begin{equation}
\Phi _{i}=\sum_{k=1}^{\dim {\cal D}\left( G_{m}\right) }\phi _{ik}\Pi _{k}
\end{equation}

where $\Pi _{k}$\ are the projectors on the eigen states basis of the ${\cal %
D}\left( G_{m}\right) $ representation and where $\phi _{ik}$\ are the
transverse coordinates of the fractional $D$ branes.

(ii) Along with the commutative $G_{m}$ symmetry, the NC algebra has an
extra automorphism $G_{m}^{aut}$ embodied by the dual group $\tilde{G}_{m}$
operating on $G_{m}$\ \ through the following adjoint action;
\begin{equation}
J_{a}\text{ }E_{b}\text{ }J_{a}^{-1}=\alpha _{ab}\text{ }E_{b},
\end{equation}
In this relation, $\left\{ E_{a}\right\} $\ and $\left\{ J_{a}\right\} $ are
respectively \ the generators of \ the ${\cal D}\left( G_{m}\right) $\ and $%
{\cal D}\left( G_{m}^{aut}\right) $\ representations and\ the complex number
$\alpha _{ab}$\ are some given group cocycles which are basically given by
the $G_{m}$ group elements.\ As such the group symmetry of the commutative
geometry will be not only given by the $G_{m}$ group, \ but rather a larger
non commutative one namely $G_{m}\ast $ $G_{m}^{aut}$.

(iii) Solutions for the $\Phi _{i}$'s can be easily worked out; one has just
to take the $\Phi _{i}$'s either in ${\cal D}\left( G_{m}\right) $ or in $%
{\cal D}\left( G_{m}^{aut}\right) $. In other words, the $\Phi _{i}$ 's are
solved either by help of the $E_{a}$ generators of ${\cal D}\left(
G_{m}\right) $ as shown here below; see eqs(4.9-10) for the quintic case,
\begin{equation}
\Phi _{i}=\phi _{i}\ \prod_{a=1}^{m}E_{a}^{p_{i}^{a}},
\end{equation}
or again by help of the $J_{a}$ generators of the automorphism symmetry $%
{\cal D}\left( G_{m}^{aut}\right) $ as; see also eqs(5.16),
\begin{equation}
\Phi _{i}=\phi _{i}\ \prod_{a=1}^{m}J_{a}^{q_{i}^{a}}.
\end{equation}

\qquad In this paper we will develop further the CP algebra method for a
large class of Calabi-Yau orbifolds with discrete torsion and too
particularly orbifolds of the Calabi-Yau homogeneous hypersurfaces ${\cal H}%
_{n}$. As a toy model, we will focus our attention on the non commutative
quintic ${\cal Q}$. There, the discrete symmetry group $G$ will be taken as $%
Z_{5}^{3}$ for ${\cal Q}$ and more generally as $Z_{n}^{n+2}$ for complex $n$
dimensional Calabi-Yau hypersurfaces. NC {\it non homogeneous} Calabi-Yau
hypersurfaces, and too particularly the so called local
Calabi-manifolds using toric methods, will be described elsewhere$[33]$

\section{Non commutative Quintic ${\cal Q}^{nc}$}

\qquad For self containment, we deserve this section to recall some general
features on the commutative quintic; then we expose the constrained method
for deriving the $Z$ realization of the non commutative one; ${\cal Q}^{nc}$.

\subsection{Commutative Quintic ${\cal Q}$}

\qquad To begin consider the complex analytic homogeneous hypersurface,

\begin{equation}
{\cal P}_{5}{(z_{1},\dots,z_{5})}%
=z_{1}^{5}+z_{2}^{5}+z_{3}^{5}+z_{4}^{5}+z_{5}^{5}+
a_{0}\prod_{i=1}^{5}z_{i}=0,
\end{equation}

where $(z_{1},z_{2},z_{3},z_{4},z_{5})$ are commuting homogeneous
coordinates of ${\bf CP}^{4}$, the complex four dimensional projective
space, and $a_{0}$ a non zero complex moduli parameter. This polynomial
describes a well known Calabi-Yau threefolds namely the commutative quintic
denoted in this paper by ${\cal Q}$. Besides invariance under permutations
of the $z_{i}$ variables, ${\cal Q}$ has a set of geometric discrete
isometries acting on the homogeneous coordinates $z_{i}$ as:

\begin{equation}
z_{i}\rightarrow z_{i}\omega ^{q_{i}^{a}},
\end{equation}

with $\omega ^{5}=1$ and $q_{i}^{a}$ integer charges to which we will refer
here below as the Calabi-Yau charges. These are the entries of the following
five dimensional vectors,

\begin{eqnarray}
{\bf q^{1}} &=&{\bf (}%
q_{1}^{1},q_{2}^{1},q_{3}^{1},-q_{1}^{1}-q_{2}^{1}-q_{3}^{1},0{\bf )}
\nonumber \\
{\bf q^{2}} &=&{\bf (}%
q_{1}^{2},q_{2}^{2},q_{3}^{2},-q_{1}^{2}-q_{2}^{2}-q_{3}^{3},0{\bf )} \\
{\bf q^{3}} &=&{\bf (}%
q_{1}^{3},q_{2}^{3},q_{3}^{3},-q_{1}^{3}-q_{2}^{3}-q_{3}^{3},0{\bf ).}
\nonumber
\end{eqnarray}

In these relations, we have set $q_{5}^{a}=0$, a useful feature which can be
thought of as corresponding to working in \ \ ${\cal U}_{\alpha }\left[
z_{1},\dots ,z_{4},z_{5}=1\right] $, \ \ the ${\cal Q}$ local coordinate
patch where $z_{5}=1$. For illustrating applications, \ we will mainly use
the following special choice $q_{1}^{a}=1,q_{2}^{1}=q_{3}^{2}=q_{4}^{3}=-1$;
all remaining ones are equal to zero. In other words;

\begin{eqnarray}
{\bf q^{1}} &=&{\bf (}1,-1,0,0,0{\bf )}  \nonumber \\
{\bf q^{2}} &=&{\bf (}1,0,-1,0,0{\bf )} \\
{\bf q^{3}} &=&{\bf (}1,0,0,-1,0{\bf ).}  \nonumber
\end{eqnarray}

The $q_{i}^{a}$ charges in eqs(3.2) and eqs(3.3-4) are defined modulo five; $%
q_{i}^{a}\equiv q_{i}^{a}+5{\bf Z} $, and satisfy naturally the identities

\begin{equation}
\sum_{i=1}^{5}q_{i}^{a}=0 \ \ modulo (5); \ a=1,2,3.
\end{equation}

These constraint eqs are necessary as required by invariance under eqs(3.2)
of the \ \ $a_{0}\prod_{i=1}^{5}z_{i}$\ \ monomials of eq(3.1); they ensure
that the holomorphic hypersurface(3.1) is a Calabi-Yau manifold. The
constraint eqs(3.5) are just the analogue of the vanishing of the first
Chern class of the quintic; $c_{1}({\cal Q})=0$,\ in differential geometry
language. These are also the conditions under which the underlying effective
field theory flows in the infrared to a CFT $[34,35].$

In section 6, we shall show that given the $q_{i}^{a}$\ charges and a
torsion matrix, we can build extra symmetries of eqs(3.1) acting as $%
z_{i}\rightarrow z_{i}\omega ^{p_{i}^{a}}$ where now the $p_{i}^{a}$\
charges are given by,
\begin{eqnarray}
p_{i}^{a} &=&\eta _{ab}\text{ }q_{i}^{b},  \nonumber \\
\sum_{i=1}^{5}p_{i}^{a} &=&0,
\end{eqnarray}

with $\eta _{ab}$\ a $3\times 3$\ matrix of \ $SL\left( 3;{\bf Z}\right) .$
The $\eta _{ab}$ matrix turns out to play an important role in our present
study; it encodes the effects of $G^{aut}\simeq {\bf Z}_{5}^{aut\text{ }3}$;
the group of automorphisms of $G$. \ We will show later that the matrix $\eta
_{ab}$\ is the carrier of the discrete torsion of the ${\bf Z}_{5}^{3}$
symmetry; its antisymmetric part $\eta _{ab}-\eta _{ba}$ is related to the
logarithm of the $t_{ab}$\ matrix of eqs(2.1).

In summary, one should retain that, if we denote by ${\cal R}_{5}\left[
[z_{1},\dots ,z_{5}\right] ]$; \ ${\cal R}_{5}$ for short, the ring of
complex holomorphic and homogeneous polynomials of degree five on ${\bf CP}%
^{4}$ and by \ $G_{\left[ \nu \right] }\subseteq G={\bf Z}_{5}^{3}$, the
subgroups of $G$ and $\ G_{\left[ \nu \right] }^{aut}\subseteq G^{aut}\simeq
{\bf Z}_{5}^{aut\text{ }3}$; the automorphisms of $G_{\left[ \nu \right] }$,
then one can build various orbifolds of the commutative quintic as shown
here below,

\begin{equation}
{\cal Q}^{\left[ \nu \right] }={\cal R}_{5}/G_{\left[ \nu \right] }.
\end{equation}

For later use, we will consider the two following examples of the quintic
orbifolds ${\cal Q}^{\left[ 1\right] }$ and ${\cal Q}^{\left[2 \right] }$
associated respectively with\ $G_{\left[ 1\right] }={\bf Z}_{5}$\ and $G_{%
\left[ 2\right] }={\bf Z}_{5}^{2}$\ subgroups of $G={\bf Z}_{5}^{3}$,
\begin{eqnarray}
{\cal Q}^{\left[ 1\right] } &=&{\cal R}_{5}/{\bf Z}_{5},  \nonumber \\
{\cal Q}^{\left[ 2\right] } &=&{\cal R}_{5}/{\bf Z}_{5}^{2}.
\end{eqnarray}

\qquad Throughout this study, we will mainly stay in the coordinate patch \
\ ${\cal U}_{(\alpha )}\left[ z_{1},\dots ,z_{4},z_{5}=1\right] $\ \ as
required by the $q_{i}^{a}$ choice made in eqs(3.4). The move to an other
patch of the quintic, say \ \ ${\cal U}_{(\beta )}\left[ w_{1},\dots
,w_{4},w_{5}=1\right] $\ \ with Calabi-Yau charges $r_{i}^{a}$, is ensured
by holomorphic transition functions $\phi _{(\alpha ,\beta
)}^{(r_{i}^{a},q_{i}^{a})}$ carrying appropriate ${\bf Z}_{5}^{3}$ charges.
Note that on the coordinate patch \ ${\cal U}_{(\alpha )}\left[ z_{1},\dots
,z_{4},z_{5}=1\right] $, the full ${\bf Z}_{5}^{3}$\ orbifold symmetry has
no fix point, the only stable one, namely $(0,0,0,0,1)$, does not belong to $%
{\cal R}_{5}/{\bf Z}_{5}^{3}$. \ However thinking of the quintic orbifold $\
{\cal R}_{5}/{\bf Z}_{5}^{3}$ as either a ${\bf Z}_{5}$ orbiofld of \ ${\cal %
R}_{5}/{\bf Z}_{5}^{2}$, that is $\ {\cal R}_{5}/{\bf Z}_{5}^{3}\sim \left(
{\cal R}_{5}/{\bf Z}_{5}^{2}\right) /{\bf Z}_{5}\sim {\cal R}_{5}$'$/{\bf Z}%
_{5}\ $or again as a \ ${\bf Z}_{5}^{2}$ \ orbifold of ${\cal R}_{5}/{\bf Z}%
_{5}$; i.e $\ {\cal R}_{5}/{\bf Z}_{5}^{3}\sim \left( {\cal R}_{5}/{\bf Z}%
_{5}\right) /{\bf Z}_{5}^{2}\sim {\cal R}_{5}$''$/{\bf Z}_{5}$,\ one can
consider the fixed points of the respective singular spaces of \ ${\cal %
R^{\prime }}_{5}/{\bf Z}_{5}$ \ and \ ${\cal R^{\prime \prime }}_{5}/{\bf Z}%
_{5}^{2}$.\ \ This procedure is also equivalent to set to zero some of the
Calabi -Yau vector charges associated with the ${\bf Z}_{5}^{3}$ symmetry.
For example the orbifold${\cal \ R}$''$_{5}/{\bf Z}_{5}$ may also be thought
of as ${\cal R}_{5}/{\bf Z}_{5}$\ by setting ${\bf q^{2}}={\bf q^{3}}={\bf 0}
$ and so may be compared to ${\cal Q}^{\left[ 1\right] }$ while by using a
similar reasoning ${\cal R}_{5} $'$/{\bf Z}_{5}$\ can be compared with $%
{\cal Q}^{\left[ 2\right] }$. \ In section7, we shall use both of the spaces
to study fractional branes and give more explicit examples.

\subsection{Non Commutative Quintic: Constrained Method}

\qquad A way to get non commutative extensions of the quintic ${\cal Q}$ by
using discrete torsions, is to start from the complex homogeneous
hypersurface (3.1) and choose a coordinate patch, say $z_{5}=1$ and so $%
q_{5}^{a}=0$. Then associate to the $\left\{ z_{1},z_{2},z_{3},z_{4}\right\}
$\ local variables, the set of $5\times 5$\ matrix operators \ $\left\{
Z_{1},Z_{2},Z_{3},Z_{4}\right\} $\ and \ $Z_{5}$\ with the identity matrix\ $%
I_{id}$. The non commutative quintic ${\cal Q}^{nc}$, associated to eqs(3.1)
and eqs(3.4) is a non commutative algebra generated by the $Z_{i}$'s. It\ is
just a special subalgebra \ of the ring of functions on the space of
matrices ${\cal M}at(5,{\bf C)}$ and reads in term of the $Z_{i}$ generators
as;
\begin{eqnarray}
Z_{i}Z_{j} &=&\beta _{ij}\ Z_{j}Z_{i}\qquad i,j=1,...,4,  \nonumber \\
Z_{i}Z_{5} &=&Z_{5}Z_{i},
\end{eqnarray}
where $\beta _{ij}$ is an invertible matrix constrained by
\begin{eqnarray}
\beta _{ji} &=&\beta _{ij}^{-1}  \nonumber \\
\beta _{j1}\beta _{j2}\beta _{j3}\beta _{j4} &=&1,\qquad \forall j,
\nonumber \\
\beta _{i1}^{5} &=&1,\qquad \forall i,  \nonumber \\
\beta _{i2}^{5} &=&1,\qquad \forall i, \\
\beta _{i3}^{5} &=&1,\qquad \forall i,  \nonumber \\
\beta _{i4}^{5} &=&1,\qquad \forall i.  \nonumber
\end{eqnarray}
These constraint eqs reflect just the property that the commutative ${\cal Q}
$ should be in the centre ${\cal Z}({\cal Q}^{nc})$ of eqs(3.7). In other
words ${\cal Q}\equiv {\cal Z(Q}^{nc})$, or equivalently;
\begin{eqnarray}
\left[ Z_{j},Z_{i}^{5}\right] &=&0,  \nonumber \\
\left[ Z_{j},\prod_{i=1}^{4}Z_{i}\right] &=&0.
\end{eqnarray}
A class of solutions for possible $\beta _{ij}$'s is obtained as follows:
First parameterize $\beta _{ij}$ as;
\begin{equation}
\beta _{ij}=\exp i\left( \frac{2\pi }{5}L_{ij}\right) =\omega ^{L_{ij}}.
\end{equation}
where $L_{ij}$\ is a $5\times 5$ antisymmetric matrix satisfying moreover
\begin{equation}
\sum_{i=1}^{5}L_{ij}=0.
\end{equation}

\qquad A careful inspection of eqs(3.11) shows that this constraint comes
from the term $a_{0}z_{1}z_{2}z_{3}z_{4}z_{5}$ \ exactly as for the
Calabi-Yau condition eqs(3.5). This means that a possible solution of (3.11)
is to take $L_{ij}$ as
\begin{equation}
L_{ij}=n_{ab}\left( q_{i}^{a}q_{j}^{b}-q_{j}^{a}q_{i}^{b}\right) \\
\equiv m_{ab}\ q_{i}^{a}q_{j}^{b},
\end{equation}
where $n_{ab}$ is a priori an arbitrary $3\times 3$ matrix of integer
entries. The matrix $L_{ij}$ is built as bi-linears of the $q_{k}^{c}$
charge vectors; ensuring automatically the condition $\sum_{i=1}^{5}L_{ij}=0$%
. This why, we shall still refer to the constraint $\sum_{i=1}^{5}L_{ij}=0$\
as the Calabi-Yau condition. Moreover, since eq(3.14) can also be written as
$L_{ij}=\ m_{ab}\ q_{i}^{a}q_{j}^{b}$, with $m_{ab}=n_{ab}-n_{ba}$, one
suspects that the $m_{ab}$ matrix should be linked to the structure
constants of the underlying $\left( {\bf Z}_{5}\ast {\bf Z}_{5}^{aut}\right)
^{3}$ symmetries. We will show, in section 6, that $L_{ij}$ should read in
fact as;
\begin{equation}
L_{ij}=\ p_{i}^{a}q_{j}^{a}-p_{j}^{a}q_{i}^{a},
\end{equation}
where $p_{i}^{a}$ are the charges of a non manifest invariance of eqs(3.1).
One may think about this latter invariance as a hidden symmetry induced by
discrete torsions and, by analogy with ${\bf Z}_{5}{}^{3}$, can be defined
as acting on the $z_{i}$ variables as;
\begin{equation}
z_{i}\rightarrow z_{i}\omega ^{p_{i}^{a}},
\end{equation}
with $p_{i}^{a}=n_{ab}q_{i}^{b}.$ We will prove moreover that the $m_{ab}$
coupling matrix is nothing but the inverse of the logarithm of the discrete
torsion matrix ${\bf t}_{ab}$ of the group $\left( {\bf Z}_{5}\ast {\bf Z}%
_{5}^{aut}\right) ^{3}$.

\qquad To have an idea on how the formulae of the constrained method work in
practice, let us go the local patch $z_{5}=1$ with $q_{i}^{a}$ charges as in
eq(3.4) and perform some explicit calculations. Setting $m_{12}=k_{1}$, $%
m_{23}=k_{2}$ and $m_{13}=k_{3}$\ are integers modulo $5$, the $L_{ij}$
matrix reads as:

\begin{equation}
L_{ij}=\left(
\begin{array}{ccccc}
0 & k_{1}-k_{3} & -k_{1}+k_{2} & k_{3}-k_{2} & 0 \\
-k_{1}+k_{3} & 0 & k_{1} & -k_{3} & 0 \\
k_{1}-k_{2} & -k_{1} & 0 & k_{2} & 0 \\
-k_{3}+k_{2} & k_{3} & -k_{2} & 0 & 0 \\
0 & 0 & 0 & 0 & 0
\end{array}
\right) .
\end{equation}

In this case, the generators of the non commutative quintic ${\cal Q}^{nc}$
satisfy the following algebra:

\begin{eqnarray}
Z_{1}Z_{2} &=&\omega ^{k_{1}-k_{3}}Z_{2}Z_{1},\qquad Z_{1}Z_{3}=\omega
^{-k_{1}+k_{2}}Z_{3}Z_{1},  \nonumber \\
Z_{1}Z_{4} &=&\omega ^{k_{3}-k_{2}}Z_{4}Z_{1},\qquad Z_{2}Z_{3}=\omega
^{k_{1}}Z_{3}Z_{2}, \\
Z_{2}Z_{4} &=&\omega ^{-k_{3}}Z_{4}Z_{2},\qquad Z_{3}Z_{4}=\omega
^{k_{2}}Z_{4}Z_{3} .  \nonumber
\end{eqnarray}

Putting $\alpha =\omega ^{k_{1}-k_{3}},$ $\beta =\omega ^{k_{1}-k_{2}}$ and $%
\gamma =\omega ^{k_{3}}$, one discovers the algebra of $[2]$ given by
eqs(1.1). Moreover the solution of these eqs read, up to a normalization
factor, as:

\begin{eqnarray}
Z_{1} &=&x_{1}{\bf P}_{\omega ^{k_{1}+k_{2}+k_{3}}}{\bf Q}^{3}  \nonumber \\
Z_{2} &=&x_{2}{\bf P}_{\varpi ^{k_{1}}}{\bf Q}^{-1}  \nonumber \\
Z_{3} &=&x_{3}{\bf P}_{\varpi ^{k_{2}}}{\bf Q}^{-1} \\
Z_{4} &=&x_{4}{\bf P}_{\varpi ^{k_{3}}}{\bf Q}^{-1},  \nonumber
\end{eqnarray}

where $x_{i}$\ are as in eqs(2.), $\varpi $ is the complex conjugate of $%
\omega $ and where

\begin{equation}
{\bf P}_{\alpha }=diag(1,\alpha,\alpha^{2},\alpha^{3},\alpha ^{4}),\quad
{\bf Q}=\left(
\begin{array}{ccccc}
0 & 0 & 0 & 0 & 1 \\
1 & 0 & 0 & 0 & 0 \\
0 & 1 & 0 & 0 & 0 \\
0 & 0 & 1 & 0 & 0 \\
0 & 0 & 0 & 1 & 0
\end{array}
\right)
\end{equation}

with $\alpha \ $standing for $\omega ^{k_{1}},\omega ^{k_{2}}$,\ $\omega
^{k_{3}}$ and their products. This solution shows clearly that $Z_{i}^{5}$, the product $\prod_{i=1}^{4}Z_{i}$\ and their linear combination are all
of them in the centre ${\cal Z}{(Q}^{nc})$ of the non commutative algebra $%
{\cal Q}^{nc}$.

\section{NC Geometry versus Discrete Torsion}

\qquad In quantum physics, non commutativity appears in different ways and
has various origins and different interpretations $\cite{36,11,12}$. At very
low energy effective field theory models such as in the Chern Simons model
of the fractional quantum Hall effect $\cite{37,38,39}$ , NC geometry is
generated by a stronger enough constant external magnetic field $B$ which
couples the electrons two dimensional position vectors $x^{i}\left( t\right) $
as \ $B\varepsilon _{ij}x^{i}\left( t\right) \frac{\partial x^{j}\left(
t\right) }{\partial t}$. Quantization rules lead to a non vanishing
commutator for the position vectors; i.e $\left[ x^{i},x^{i}\right] =i\frac{\epsilon^{ij}}{B\nu_L}$, with $\nu_L=\frac{1}{k}$ being the Laughlin filling fraction. For stron $B$, the quantum properties of the system of electrons are described by a non commutative Chern-Simons gauge theory on a D2 brane. Such a 2D condensed matter phase  has recieved recently an important interest due to similarities with branes systems in superstring theories.  At very
high energies, say around the Plank scale as in string theory, non
commutativity is generated by the antisymmetric NS-NS $B_{\mu \nu }$ field
and it is linked to the existence of open strings ending on $D$ branes\ and
a dynamics governed by a boundary conformal invariance $\cite{12}$. This
issue has been subject to much interest during the few last years in
connection with the derivation of non commutative solitons and the study of
the tachyon condensation by following the GMS method. In M theory, non
commutative geometry comes as a non trivial solution for the matrix model
compactification, but still generated by an antisymmetric field; the eleven
dimensional three form gauge field $C_{\mu \nu \rho }$\cite{11}. Elements $%
g_{1}$ and $g_{1}$ of the group of automorphism symmetries of the matrix
model on a two torus ${\cal T}^{2}$ are in general governed by the central
relation $g_{1}g_{2}g_{1}^{-1}g_{2}^{-1}$ taken to be proportional
to the identity operator;\  that is $g_{1}g_{2}g_{1}^{-1}g_{2}^{-1}=\lambda $ $%
I_{id}$, with $\lambda \in {\bf C}^{\ast }$  as required by the Schur lemma
\cite{11,14}.
\qquad  In quantum mathematics, NC geometry is viewed as an algebraic structure ${\cal M%
}_{\hbar }\left[ X_{1},\dots ,X_{N}\right] $ going beyond the usual ${\cal C}%
\left[ x_{1},\dots ,x_{N}\right] $ commutative one, with the ideal $\left\{
x_{i}x_{j}=x_{j}x_{i}\right\} $. Formally, the generic commutation relations
of the generators of the quantum algebra ${\cal M}_{\hbar }\left[
X_{1},\dots ,X_{N}\right] $, may be written as $\cite{40,41}$:

\begin{equation}
X_{I}\ast X_{J}=r_{IJ}^{KL}(X)\ \ X_{K}\ast X_{L}+\ \ b_{IJ}(x),
\end{equation}

where $r_{IJ}^{KL}(X)$ and $b_{IJ}(X)$ are some polynomials in $X_{I}$ which
may be thought of as;
\begin{eqnarray}
\ r_{IJ}^{KL}(X)\ &=&\delta _{I}^{L}\delta _{J}^{K}+\hbar \ r_{IJ}^{\prime
KL}+...,  \nonumber \\
\ \ b_{IJ}(X)\ &=&\hbar \left( \Omega _{IJ}+f_{IJ}^{K}\ X_{K}+...\right)
\end{eqnarray}
\ In the limit $\hbar \ \rightarrow 0$, \ \ $r_{IJ}^{KL}(X)\ \rightarrow
\delta _{I}^{L}\delta _{J}^{K},\ \ b_{IJ}(X)\ \rightarrow 0$;\ and so one
recovers the usual commutative structure of ${\cal C}\left[ x_{1},\dots
,x_{N}\right] .$ \ For the general cases, such for instance\ $%
(r_{IJ}^{KL}(x),b_{IJ}(x))$ \ equals to $(0,B_{IJ})$, \ $(0,f_{IJ}^{K}\
X_{K})$ and $(R_{IJ}^{KL}\ ,0)$, one gets respectively the canonical
commutator, the Lie algebra bracket and the quantum Yang-Baxter spaces $\cite
{36}$.

\qquad The non commutative structure we are dealing with in the present
paper corresponds to an other special situation where $b_{IJ}(x)=0$\ and,
\begin{equation}
r_{IJ}^{KL}(x)=\beta _{KL}\ \delta _{I}^{L}\ \delta _{J}^{K};,
\end{equation}
with \ $\beta _{IJ}$ is a root of unity. This NC geometry is generated by
discrete torsion matrix of the orbifold group ${\bf Z}_{5}^{3}$ and has much
to do with NC Fuzzy torii representations. Since discrete torsions are
involved in string compactifications on orbifolds and twisted string
sectors, one expects that such NC structure could play some role in string
theory on orbifolds. As we shall show by explicit analysis in this paper,
see sections 4 and 7, NC geometry versus discrete torsion leads to
fractional $D$ branes at orbifold singularities and offers a way to resolve
non geometric singularities. Points of the usual geometry are replaced
by polygons in the non commutative case.

\qquad To get the right link between the non commutative quintic solutions
given in the previous section and the discrete torsions of ${\bf Z_{5}}^{3}$%
, we start by reconsidering the solutions eqs(3.19) \ by first showing that
they are not so general as claimed in $\cite{24}$. Then we give some results
regarding \ ${\bf Z}_{5}^{3}$\ \ group theoretical representations and
present a realization for the case where discrete torsions are non trivial.
\ Next, we study the ${{\bf Z}_{5}^{aut}}^{3}$ \ hidden symmetry of the
quintic raised in the previous sections and acting as in eqs(3.16).

\subsection{More on Solutions (3.19)}

\qquad A careful inspection of eqs(3.19), shows that this solution involves
various group elements of the five dimensional representation of ${\bf Z}%
_{5}\ast {\bf Z}_{5}^{aut}$; \ these are ${\bf P}_{\omega ^{k_{1}}},{\bf P}%
_{\omega ^{k_{2}}}$ and ${\bf P}_{\omega ^{k_{3}}}$ and powers of ${\bf Q}$.
Since ${\bf P}_{\alpha }{\bf P}_{\beta }={\bf P}_{\alpha \beta }$, it
follows that these solutions may also be written as
\begin{eqnarray}
Z_{1} &=&x_{1}{\bf P}^{k_{1}+k_{2}+k_{3}}{\bf Q}^{3}  \nonumber \\
Z_{2} &=&x_{2}{\bf P}^{-k_{1}}{\bf Q}^{-1}  \nonumber \\
Z_{3} &=&x_{3}{\bf P}^{-k_{2}}{\bf Q}^{-1} \\
Z_{4} &=&x_{4}{\bf P}^{-k_{3}}{\bf Q}^{-1},  \nonumber
\end{eqnarray}
where we have set ${\bf P}_{\omega }={\bf P}$ \ The symmetry of the
solutions (3.19) is then ${\bf Z}_{5}\ \ast {\bf Z}_{5}^{aut}$ generated by $%
{\bf P}$ and ${\bf Q}$ satisfying ${\bf Q\ P\ Q}^{-1}=\omega \ {\bf P}.$
Comparing this relation with eq(2.3), i.e \ $Z_{i}=x_{i}\
\prod_{a}T_{\{a_{i}\}}$, one may identify easily the various elements of $\
{\bf Z}_{5}\ \ast {\bf Z}_{5}^{aut}$ entering in the building of the
solutions. \ For the operators system $\left\{ {\bf P}_{\omega ^{k_{1}}},%
{\bf P}_{\omega ^{k_{2}}},{\bf P}_{\omega ^{k_{3}}},{\bf Q}\right\} $
appearing in the building of $Z_{2},$ $Z_{3}$ and $Z_{4}$, we have the
following torsion matrix;

\begin{equation}
{\bf t}_{\mu \nu }=\left(
\begin{array}{cccc}
1 & 1 & 1 & \omega ^{k_{1}} \\
1 & 1 & 1 & \omega ^{k_{2}} \\
1 & 1 & 1 & \omega ^{k_{a}} \\
\omega ^{-k_{1}} & \omega ^{-k_{2}} & \omega ^{-k_{3}} & 1
\end{array}
\right) .
\end{equation}

Setting $\theta _{\mu \nu }=\frac{-5i}{2\pi }\log {\bf t}_{\mu \nu },$ we
can rewrite the above relation as
\begin{equation}
{\bf t}_{\mu \nu }=\left(
\begin{array}{cccc}
0 & 0 & 0 & k_{1} \\
0 & 0 & 0 & k_{2} \\
0 & 0 & 0 & k_{3} \\
-k_{1} & -k_{2} & -k_{3} & 0
\end{array}
\right) ;
\end{equation}

showing clearly that discrete torsion exist whenever one of the $k_{a}$
integers are non zero; i.e $\ k_{a}\neq 0$. One of the remarks concerning
eqs(3.19) is that $\left( {\bf Z}_{5}\ \ast {\bf Z}_{5}^{aut}\right) ^{3}$\
invariance is not fully apparent. We will turn to this feature later on when
we derive the general solutions later on; but for the moment let us make a
remark concerning the NC solutions one gets in general.

\qquad Due to discrete torsions, the algebraic structure of the $D$ branes
wrapping the compact manifold change. Brane points $\{z_{i}\}$ of
commutative geometry become, in presence of torsion, fibers based on $%
\{z_{i}\}$ and valued in the algebra ${\cal A}_{G}$ of the torsion group $G$%
. As such, the correspondence;

\begin{equation}
z_{i}.1=\ \rightarrow \ \ Z_{i}=\ \ \ \sum_{k,l=1}^{5}\ Z_{i}^{kl}\ |k><l|.
\end{equation}
has a nice interpretation in terms of quiver diagrams. Associating to each \
${\bf e}_{kl}\equiv |k><l|$\ matrix vector basis, a straight line ( string)
oriented from the $k$ end to the $l$ one and to each ${\bf e}_{kk}={\pi_{k}}%
\equiv |k><k|$ projector, a loop starting and ending at the position $k,$ as
shown on the following table, one may draw a quiver diagram for each $Z_{i}$
matrix generator of the NC algebra.

\begin{equation}
\begin{tabular}{|l|l|}
\hline
Operators & \ \ \ \ diagrams \\ \hline
${\bf a}_{k}^{+}\equiv |k><k+1|$ & $\ \ \ \ _{k}\quad \bullet $ $%
\longrightarrow \bullet \quad _{\left( k+1\right) }$ \\ \hline
${\bf a}_{k}^{-}\equiv |k+1><k|$ & $\ \ \ _{\left( k+1\right) }\quad \bullet
$ $\longleftarrow \bullet \quad _{k}$ \\ \hline
$\prod_{j=0}^{n}{\bf a}_{k+j}^{+}\equiv |k><k+n|$ & $\ \ \ \ _{k}\quad
\bullet \longrightarrow \bullet \quad _{\left( k+n\right) }$ \\ \hline
$\prod_{j=0}^{n}{\bf a}_{k+j}^{-}\equiv |k+n><k|$ & $\ \ \ _{\left(
k+n\right) }\quad \bullet \longleftarrow \bullet \quad _{k}$ \\ \hline
$\pi _{k}={\bf a}_{k}^{+}{\bf a}_{k}^{-}=|k><k|$ & $\ \ \ \ _{k}\bullet $ $%
\leftrightharpoons $ $\bullet $ $\ _{\left( k+1\right) }\quad \equiv \
_{k}\bigcirc \quad \equiv \bullet $ \\ \hline
$\pi _{k}={\bf a}_{k-1}^{-}{\bf a}_{k-1}^{+}=|k><k|$ & $\ \ \ \ _{\left(
k-1\right) }\bullet $ $\leftrightharpoons $ $\bullet $ $\ _{k}\quad \equiv \
\bigcirc $ $_{k}\quad \equiv \bullet $ \\ \hline
\end{tabular}
\end{equation}
\bigskip

\bigskip

\qquad Using these rules, one sees that for the quintic the generic quiver
diagram for the $Z_{i}$\ operators is given by a polygon with five
vertices (a pentagon) and in general twenty links joining the various
vertices; see figure 1. Since\ a vector basis type ${\bf e}_{k\left(
k+n\right) }=|k><k+n|$ can be usually decomposed as the product of $n$ basis
links as shown on this equation
\[
{\bf e}_{k\left( k+n\right) }={\bf a}_{k}^{+}{\bf a}_{k+1}^{+}...{\bf a}%
_{k+n}^{+},
\]
where ${\bf a}_{k}^{+}={\bf e}_{k\left( k+1\right) }$, one concludes that
points in NC geometry are roughly speaking described by pentagons.
\begin{figure}[tbh]
\begin{center}
\epsfig{file=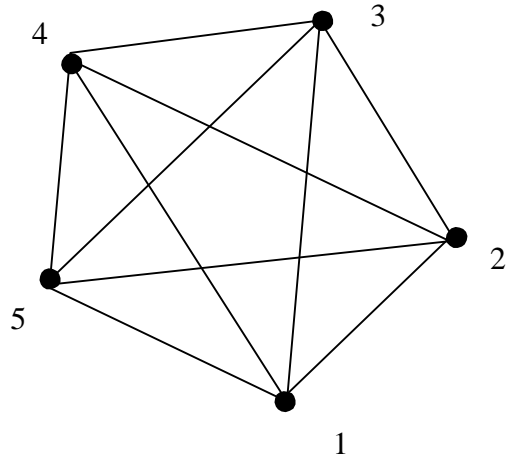}
\end{center}
\caption{{\protect\small {\it A pentagon diagram representing points of the
NC quintic, ${\cal Q} ^{nc}$. Since generic links may usually be decomposed
in terms of the $|j><j+1|$ edges, this quiver diagram can also be thought as
a polygon of five vertices and so NC points of ${\cal Q} ^{nc}$ may be
imagined as pentagons. }}}
\end{figure}
\bigskip

We will be use later on these diagrams to study the the \ fractional $D$
brane at singularities of the orbifold of the quintic.

\subsection{Discrete Torsion Matrix of ${\bf Z}_{5}^{3}$}

\qquad Here we want to study general ${\cal D}\left( G\right) $
representations of the ${\bf Z}_{5}^{3}$ in presence of torsion. We shall
first consider the simplest situation where all ${\bf \ Z}_{5}$ factors
commute amongst themselves; then we discuss the case where they do not
commute.

\subsubsection{Free torsion case}

\qquad Naively, the geometric ${\bf Z}_{5}^{3}$ can be seen as the product
of three abelian ${\bf Z}_{5}$ group factors whose generators may be
defined, by help of the tensor product, as follows:

\begin{eqnarray}
E_{1}&=& P_{1}\otimes I_{id}\otimes I_{id},  \nonumber \\
E_{2}&=& I_{id}\otimes P_{2}\otimes I_{id}, \\
E_{3}&=&I_{id}\otimes I_{id}\otimes P_{3} ,  \nonumber
\end{eqnarray}

The $E_{a}^{\prime }$ s are the generators of the three \ ${\bf Z}_{5}$\
factors of \ ${\bf Z}_{5}^{3}$\ ; they satisfy the cyclic property $%
E_{a}^{5}={\bf I}_{{\cal D}\left( G\right) }$ \ following from the
individual identities $P_{a}^{5}=I_{id}$. Since they are commuting
operators; i.e,

\begin{equation}
{E_{a}}.{E_{b}}={E_{b}}.{E_{a}},
\end{equation}

they can be diagonalized simultaneously in the same basis \ $\{|a,i>;1\leq
a\leq 3;\ 1\leq i\leq 5\}$. As such the $P_{a}^{\prime }$ s can be thought
more a less as in eq(3.20) and $E_{a}^{\prime }$ s as blocks of diagonal
matrices. Using the convention notations ${\bf a}_{n_{a}}^{+}\ =\
|a,n>\otimes <a,n+1|,$ $\ {\bf a}_{n_{a}}^{-}\ =\ |a,n+1>\otimes <a,n|$ and $%
\ \pi _{n_{a}}\ =\ {\bf a}_{n_{a}}^{+}\ {\bf a}_{n_{a}}^{-},$ and the
graphic representation eqs(4.8), we can also represent the following
operators decompositions;

\begin{eqnarray}
I_{id}\ &=&\ \sum_{n=1}^{5}\ \ \ \pi _{n},\quad P_{a}\ =\ \sum_{n=1}^{5}\
\alpha _{a,n}\ \ \pi _{n},  \nonumber \\
Q_{a}\ &=&\ \sum_{n=1}^{5}\ {\bf a}_{\ a,n}^{+};\quad
Q_{a}^{-1}=\sum_{n=1}^{5}{\bf a}_{\ a,n}^{-},
\end{eqnarray}

in terms of quiver diagrams as shown on figures 2. For explicit
computations, we will drop the index a by working in special matrix
realizations.
\begin{figure}[tbh]
\begin{center}
\epsfxsize=12cm \epsffile{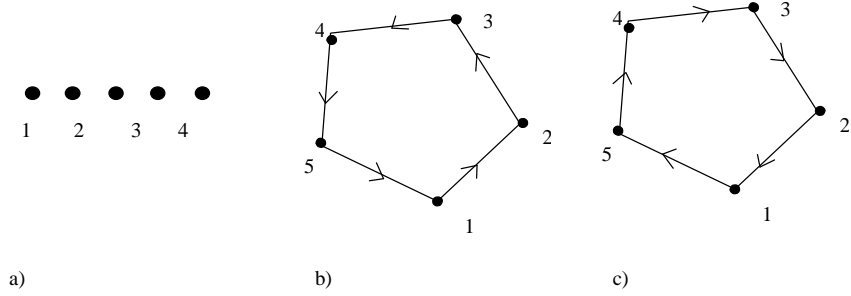}
\end{center}
\caption{{\protect\small {\it The completely reducible diagram Fig2.a
represents the identity operator of the five dimensional representation $%
{\cal D}({\bf Z}_{5})$. Fig2.b is an oriented pentagon representing the $Q$
automorphism operators while fig2.c its inverse. }}}
\end{figure}

\qquad Generic elements $g$ of \ ${\bf Z}_{5}^{3}$\ split also as \ $%
g_{1}\otimes g_{2}\otimes g_{3}$, \ the tensor product of the $g_{a}$ \ 
elements of the ${\bf Z}_{5}^{\prime }$ s. Since ${\bf Z}_{5}^{3}$\ \ is a
cyclic group, one can expand these the $g^{\prime }$ s,\ in terms of the $%
E_{a}$ generators as follows:

\begin{equation}
g=\prod_{i_{1},i_{2},i_{3}=1}^{5}\ \gamma
_{i_{1}i_{2}i_{3}}E_{1}^{i_{1}}E_{2}^{i_{2}}E_{3}^{i_{3}},
\end{equation}

where the $\gamma _{i_{1}i_{2}i_{3}}$ coefficients are such that\ $\ {\gamma
_{i_{1}i_{2}i_{3}}}^{5}=1$. Indeed, since the group multiplication law on\ $%
{\bf Z}_{5}^{3}$\ is defined as usual by performing multiplications of
individual elements; that is $g.g^{\prime }=g_{1}.g_{1}^{\prime }\otimes
g_{2}.g_{2}^{\prime }\otimes g_{3}.g_{3}^{\prime }$; \ we have ${E_{a}}^{5}=%
{\bf I}_{{\cal D}\left( {\bf Z}_{5}^{3}\right) }$, with ${\bf I}_{{\cal D}%
\left( {\bf Z}_{5}^{3}\right) }$ standing for $I_{{\cal D}\left( {\bf Z}%
_{5}\right) }\otimes I_{{\cal D}\text{'}\left( {\bf Z}_{5}\right) }\otimes
I_{{\cal D}\text{''}\left( {\bf Z}_{5}\right) }\equiv I_{id}\otimes
I_{id}\otimes I_{id}$ \ group representation identity.

As all elements of \ ${\bf Z}_{5}^{3}$\ can be expressed as powers of ${E_{a}%
}^{\prime }$s, we will focus our attention on these generators as well as on
their monomials $\{E_{{1}}^{n_{1}}E_{2}^{n_{2}}E_{3}^{n_{3}};\ 1\leq
n_{a}\leq 5\}$. The latter generate the dimension $d_{1}d_{2}d_{3}$ \ group
algebra \ ${\cal A}_{G}$\ of \ ${\bf Z}_{5}^{3}.$ The $d_{i}$\ integers are
the dimensions of the three ${\cal D}\left( {\bf Z}_{5}\right) $ factors of $%
{\cal D}\left( {\bf Z}_{5}^{3}\right) .$

\subsubsection{ Discrete Torsion}

\qquad Though abelian, the three ${\bf Z}_{5}$ factors of \ ${\bf Z}_{5}^{3}$%
\ may also not commute among each others and this situation corresponds to
the case where there exist discrete torsions. In other words given a couple
of elements $\left( g_{1},g_{2}\right) \in {\bf Z}_{5}\otimes {\bf Z}_{5},$
then $g_{1}g_{2}\neq g_{2}g_{1}$.In practice, we shall have $%
g_{1}g_{2}=\alpha $ $g_{2}g_{1}$ with $\alpha $\ is a\ fifth root a unity.
This situation, which is quite similar to the Fuzzy torii realizations known also as  the
rational non commutative torii, may be described by extending the relations
(4.7) by allowing twisting co-cycles $t_{ab}$ \ between the three ${\bf Z}%
_{5}^{3}$ factors as shown here below,

\begin{equation}
{F_{a}}\ {F_{b}}=t_{ab}\ {F_{b}}\ {F_{a}},
\end{equation}

where $t{(F_{a},F_{b})}\equiv t_{ab}$, is the torsion matrix and where the $%
F_{a}^{\prime }$ s are the new \ ${\bf Z}_{5}^{3}$\ generators; they are
essentially the $E_{a}$ 's \ times some $T_{a}$ matrices carrying the
torsion effects; $F_{a}=T_{a}$ $E_{a}$. \ Moreover, like the $E_{a}$ 's and
\ $T_{a}$ 's, the $F_{a}$ 's  satisfy equally $F_{a}^{5}={\bf I}_{{\cal D}\left( {\bf Z}%
_{5}^{3}\right) }$. From the relations (4.10), one learns several
informations on their representations:

\paragraph{\it Structure of the torsion matrix} \par
\qquad Using the cyclic group property $F_{a}^{5}={\bf I}_{{\cal D}\left(
{\bf Z}_{5}^{3}\right) }$, one sees that the torsion matrix $\ t_{ab}$ \
should be like $t_{ab}=\omega ^{\theta _{ab}}$ where $\theta _{ab}=-\theta
_{ba}$ \ is antisymmetric $3\times 3$ matrix. For $\theta =0$, one gets just
the relations (4.7). A tricky parameterization of torsion is to take $\theta
_{ab}$ \ as $\theta _{ab}=\eta _{ab}-\eta _{ba},$ where $\eta _{ab}$ \ is an $%
SL(3,{\bf Z})$ matrix, and rewrite eqs(4.10) as\

\begin{equation}
\omega ^{\eta _{ba}}\ {F_{a}}\ {F_{b}}=\omega ^{\eta _{ab}}\ {F_{b}}\ {F_{a}}.
\end{equation}

Now using the change $F_{a}=T_{a}$ $E_{a}$, one sees that the $T_{a}$\
operators are just the generators of automorphisms of \ ${\bf Z}_{5}^{3}$ as
shown on the following relation
\begin{equation}
T{_{b}}\ E{_{a}}\ {T_{b}^{-1}=}\omega ^{\eta _{ab}}\text{ }E{_{a}}
\end{equation}

Actually this relation should be compared with eqs(2.18) and (4.19-20).
Therefore the $T{_{a}}$\ 's are just the generators of ${\bf Z}_{5}^{aut%
\text{ }3}$, \ the automorphism group of \ ${\bf Z}_{5}^{3}$\ symmetry.

\paragraph{${\cal D}\left( G\right) $ \ {\it Representations}}\par

\qquad Taking the determinant of eqs(4.14), one discovers that the dimension
$d$ of the ${\cal D}\left( G\right) $ matrix representations splits as the
sum of three terms as given herebelow;

\begin{equation}
d \ = \ d_{1}+\ d_{2}+\ d_{3} \ \ \in 5{\bf Z},
\end{equation}

where \ $(d_{1},\ d_{2},\ d_{3})$\ are respectively the dimensions of the \ $%
{\cal D}\left( {\bf Z}_{5}\right) \otimes {\cal D}$'$\left( {\bf Z}%
_{5}\right) \otimes {\cal D}$''$\left( {\bf Z}_{5}\right) $\
representations. From this decomposition, one sees that the ${F_{a}}^{\prime
}$s may be realized in various ways in terms of tensor products of $%
d_{a}\times d_{a}$ matrices. A convenient parameterization of the $F_{a}$
generators, which we will use to establish our results, is given by the
following;

\begin{eqnarray}
F_{1}&=& P_{1}\otimes Q_{2}^{\eta_{12}}\otimes Q_{3}^{\eta_{13}},  \nonumber
\\
F_{2}&=&Q_{1}^{\eta_{21}}\otimes P_{2}\otimes Q_{3}^{\eta_{23}}, \\
F_{3}&=&Q_{1}^{\eta_{31}}\otimes Q_{2}^{\eta_{32}}\otimes P_{3} ,  \nonumber
\end{eqnarray}

where the $P_{a}^{\prime }$ s and the $Q_{a}^{\prime }$ s \ are as in
eq(3.19) \ with \  ${P_{a}}{Q_{b}}=\alpha _{a}\ {Q_{b}}{P_{a}}\delta _{ab}$%
,\ and where the $\eta _{ab}$'s are integers. The latters are in fact some
given integer numbers carrying torsion effect; they are the same as in
eq(4.15). Viewed as a matrix, \ $\eta _{ab}$\ should be in $SL(3,{\bf Z})$;
\ $\eta _{ab}\ \in SL(3,{\bf Z})$;\ \ and moreover should have a non
vanishing antisymmetric part;
\begin{eqnarray}
\eta _{ab}-\eta _{ba}\ &\neq &\ 0,\quad \eta _{ab}.\eta _{cb}^{-1}=\delta
_{ac},  \nonumber \\
\eta _{11}\eta _{22}\eta _{33}+\eta _{12}\eta _{13}\eta _{31}+\eta _{13}\eta
_{11}\eta _{22} &\neq &\eta _{11}\eta _{23}\eta _{32}+\eta _{12}\eta
_{11}\eta _{33}+\eta _{13}\eta _{12}\eta _{21}.
\end{eqnarray}

To fix the ideas, we set for simplicity $\alpha _{1}=\alpha _{2}=\alpha
_{3}=\omega $ and consider the special case where the three basis of the three
representation factors of ${\cal D}\left( G\right) $, namely $\{|a,i>\
;1\leq i\leq d_{a};\ \,1\leq a\leq 3\}$, have the same dimension; i.e , $%
d_{1}=d_{2}=d_{3}$ and so $d=d_{1}+d_{2}+d_{3}=3d_{1}$. Extensions to general cases are  straightforward and some of their  aspects will be given in the
conclusion. In this case and for $d_{1}=5$, the ${\bf P}_{1}(\omega )={\bf P}%
_{2}(\omega )={\bf P}_{3}(\omega )={\bf P}$ \ and $\ {\bf Q}_{1}{\bf =Q}_{2}%
{\bf =Q}_{3}{\bf =Q}$, are as in eq(3.19) . With this choice, the relations $%
{\bf P}_{a}{\bf Q}{_{b}}=\alpha _{a}\ {\bf Q}_{b}{\bf P}{_{a}}\delta _{ab}$
reduce to tree identical equations of type;

\begin{eqnarray}
{\bf P.Q}\ &=&\ \omega \ {\bf Q.P},  \nonumber \\
{\bf P}^{5}\ &=&\ \ I_{id},\quad {\bf Q}^{5}\ =\ \ I_{id}.
\end{eqnarray}
Before going ahead, we want to give three  remarks and make some comments
regarding eqs(4.17) and (4.18-19). The  remarks regard the
parameterization (4.17), the automorphism symmetry and the ${\cal D}\left(
G\right) $ representations.

\paragraph{{\it The }$F_{a}$ Basis}\par

\qquad The parameterization eq(4.17) is very special, first because it is
related to the abelian one as
\begin{equation}
F_{a}\ =\ T_{a}\ E_{a},
\end{equation}
where the $T_{a}$ 's are operators built out of monomials of the following
generators;
\begin{eqnarray}
J_{1}\ &=&\ Q_{1}\otimes I_{id}\otimes I_{id},  \nonumber \\
J_{2}\ &=&\ I_{id}\otimes Q_{2}\otimes I_{id},  \nonumber \\
J_{3}\ &=&\ I_{id}\otimes I_{id}\otimes Q_{3}, \\
J_{a}^{5}\ &=&\ \ {\bf I}_{{\cal D}\left( G\right) }.  \nonumber
\end{eqnarray}
Indeed, comparing eq(4.20) with eqs(4.17), it is not difficult to check that
the $T_{a}$ 's read as follows;

\begin{eqnarray}
T_{1}\ &=&\ \ J_{2}^{\eta _{12}}.J_{3}^{\eta _{13}},  \nonumber \\
T_{2}\ &=&\ \ J_{1}^{\eta _{21}}.J_{3}^{\eta _{23}},  \nonumber \\
T_{3}\ &=&\ \ J_{1}^{\eta _{31}}.J_{2}^{\eta _{32}}, \\
T_{a}^{5}\ &=&\ \ {\bf I}_{{\cal D}\left( G\right) }.  \nonumber
\end{eqnarray}

This means that torsion will be generated by the $J_{a}$ operators. The
second thing we want to mention is that there are more general
parameterizations $F_{a}^{\prime }$ for the group generators going beyond
eqs(4.17). They extend the $F_{a}$'s \ and involve $E_{a}$'s and $J_{a}$'s
on equal footing as well as two kind of torsion matrices $n_{ab}$ and $\nu
_{ab}$\ as shown here below;

\begin{eqnarray}
F^{\prime}_{1}&=& P_{1}^{n_{11}}Q_{1}^{\nu_{11}}\otimes P_{2}^{n_{12}}
Q_{2}^{\nu_{12}}\otimes P_{3}^{n_{13}}Q_{3}^{\nu_{13}},  \nonumber \\
F^{\prime}_{2}&=&P_{1}^{n_{21}}Q_{1}^{\eta_{21}}\otimes
P_{2}^{n_{22}}Q_{2}^{\nu_{22}}\otimes P_{3}^{n_{23}}Q_{3}^{\nu_{23}}, \\
F^{\prime}_{3}&=&P_{1}^{n_{31}}Q_{1}^{\nu_{31}}\otimes P_{2}^{n_{32}}
Q_{2}^{\nu_{32}}\otimes P_{3}^{n_{32}}Q_{3}^{\nu_{33}} ,  \nonumber
\end{eqnarray}
or equivalently in terms of powers of the $E_{a}$ 's and $J_{a}$ 's;

\begin{equation}
F_{a}^{\prime }=\ \prod_{b=1}^{3}\ \left( E_{1}^{n_{ab}}J_{1}^{\nu
_{ab}}E_{2}^{n_{ab}}J_{2}^{\nu _{ab}}E_{3}^{n_{ab}}Q_{3}^{\nu _{ab}}\right) ,
\end{equation}

where $n_{ab}$ and $\nu _{ab}$ are integers modulo 5. Eqs(4.23) can be
related as well to the abelian one (4.10); but also to the parameterization
eqs(4.17) as

\begin{equation}
F^{\prime}_{a}\ =\ T^{\prime}_{a}\ \ F_{a},
\end{equation}

where $T_{a}^{\prime }$ 's are now monomials of the primitive $E_{a}$ and $%
J_{a}$ generators; i.e \ $T_{a}=T_{a}(E_{1},E_{2},E_{3};J_{1},J_{2},J_{3})$
with $T_{a}^{\prime 5}={\bf I}_{{\cal D}\left( G\right) }$. Due to this
arbitrariness in the choice of the basis of generators of the ${\cal D}%
\left( G\right) $ group representation, one learns already at this level
that there are different ways to realize the non commutative extension of
the quintic. The so called $Z$ and $\Phi $ realizations described in section
2 are just two special ones.

\paragraph{{\it More on Quantum Symmetries}}\par

\qquad The $J_{a}$ operators have a nice interpretation; they are the
primitive generators of \ the ${\bf Z}_{\ 5}^{aut}{}^{\otimes 3}$
automorphism group of the {${\bf Z}_{5}^{3}$} symmetry of the quintic. This
is clearly seen on eqs(4.24), which associate to each symmetry generator $%
E_{a}$, an automorphism $\ J_{a}$. Eqs(4.23-24) tell us that $E_{a}$ and $%
J_{a}$ generate three copies of non commutative two dimensional fuzzy torii
of parameters $\alpha _{a}$; \ ${\cal T}_{\alpha _{1}}\otimes {\cal T}%
_{\alpha _{2}}\otimes {\cal T}_{\alpha _{3}}$; and moreover show that{\ the
full symmetry of the quintic is in fact \ }$\left( {{\bf Z}_{5}\ast {\bf Z}%
_{\ 5}^{aut}}\right) ${$^{3}$.  The latter acts on the }${z}$ \ complex
variables as as in eqs(3.2) and (3.15)\newline
Furthermore, the analysis we have been developing here for discrete
torsion is valid for any complex $d$ dimension Calabi-Yau orbifolds ${\cal O}%
_{d}$ with discrete group \ ${\bf Z}_{d+2}^{\otimes d}$ generating a real $%
2d $ dimensional NC torus of type ${\cal T}_{\alpha _{1}}\otimes ...\otimes
{\cal T}_{\alpha _{d}}$. \ It is also suspected to apply as well to the
continuous limit ${\bf Z}_{\ \infty }\simeq U(1)$ where we expect the
appearance of irrational representations of non commutative Torii and
Powers-Riffel projectors [25-28] together with a continuous fractional $D$
-brane spectrum. Further details on the last issue will be reported in [33]

\paragraph{\it {Representations}}\par

\qquad The constraint eq(4.16) on the dimensions of the $F_{a}$
representations shows that it is possible to build different realizations
for the $F_{a}$ generators. The representation eqs(3.19) built in $\cite{24}$,  corresponds to take $(d_{1},d_{2},d_{3})$ equal to either $(5,0,0),\ (0,5,0)$, or $%
\ (0,0,5)$ respectively obtained by setting $\eta _{a2}=\eta _{a3}=0,\ \eta
_{a1}=\eta _{a3}=0$ \ and $\eta _{a1}=\eta _{a2}=0$. Here we shall give new
solutions of type $(d_{1},d_{2},d_{3})$ by keeping all $\eta _{ab}\neq 0$.

Now, using the relations(4.17), one can compute the expression of the $%
t_{ab} $ cocycles between the $F_{a}$ generators. We find

\begin{equation}
t{(F_{a},F_{b})}=\omega ^{(\eta _{ab}-\eta _{ba})},
\end{equation}

showing that $\theta _{ab}$ is equal to $(\eta _{ab}-\eta _{ba})$. As we
will handle much more \ $\theta _{ab}$\ and \ $\eta _{ab}$ \ than the $t{%
(F_{a},F_{b})}$ \ matrix, we will also refer to $\theta _{ab}$ and \ $\eta
_{ab}$ \ as discrete torsion matrices of the group ${\bf Z}_{5}^{3}$; \
though they are proportional to its logarithm; i.e \ \ $\theta _{ab}={\frac{%
-5i}{1+5k}}\ Log(t_{ab});\ k\in {\bf Z}$. \ Note in passing that torsion is
carried by the antisymmetric part of the matrix $\eta _{ab}$. We will also
see that the antisymmetric part of its inverse $\eta _{\left[ ab\right]
}^{-1} $ plays equally a crucial role in working out the solutions for the
NC quintic.

\subsubsection{Quiver Diagrams}

\qquad Like for the case of one abelian factor, one can also build the
projectors for full the \ ${\bf Z}_{5}^{3}$ \ group. Using the individual $%
{\bf Z}_{5}$ \ projectors $\pi _{k_{a}}={\frac{1}{5}}\sum_{i=1}^{5}\ {\omega
^{(-k_{a})}}\ {\bf P}_{\omega }^{i}$, \ we can construct a variety of
projectors on the representation space of ${\bf Z}_{5}^{3}$. First, the${\
\Pi }_{k_{a}}$ projectors on the ${\bf Z}_{5}$ representation spaces;
\begin{eqnarray}
{\ \Pi }_{k_{1}}\ &=&\ {\pi _{k_{1}}}\otimes I_{id}\otimes I_{id},  \nonumber
\\
{\ \Pi }_{k_{2}}\ &=&\ I_{id}\otimes \pi _{k_{2}}\otimes I_{id}, \\
{\ \Pi }_{k_{3}}\ &=&\ I_{id}\otimes I_{id}\otimes \pi _{k_{3}},  \nonumber
\end{eqnarray}
They have quiver diagrams more a less similar to that of ${\pi _{k_{a}},}$
except that now we have some richness coming from the decomposition of the extra
identity operators. In any way, the full quiver diagram is given by the
cross product of the individual graphs.

Second, the ${\ \Pi }_{(k_{a},k_{b})}$ \  and \  ${\ \Pi }_{(k_{1},k_{2},k_{3})}$ \ projectors on the ${\bf Z}_{5}^{2}$ \ and ${\bf Z}_{5}^{3}$ \
representation spaces, respectively obtained by taking products of ${\
\Pi }_{k_{a}}^{\prime }s$;
\begin{eqnarray}
{\ \Pi }_{(k_{a},k_{b})}\ &=&\ {\ \Pi }_{k_{a}}{\ \Pi }_{k_{b}}  \nonumber \\
{\ \Pi }_{(k_{1},k_{2},k_{3})}\ &=&\ {\ \Pi }_{k_{1}}{\ \Pi }_{k_{2}}{\ \Pi }%
_{k_{3}}.
\end{eqnarray}
Accordingly, the identity matrix ${\bf I}_{{\cal D}\left( G\right) }$
can be decomposed in different ways as shown herebelow;
\begin{equation}
{\bf I}_{{\cal D}\left( G\right) }\ =\ \sum_{k_{a}=1}^{5}\ {\ \Pi }%
_{k_{a}}=\ \sum_{k_{a},k_{b}=1}^{5}\ {\ \Pi }_{k_{a},k_{b}}\
=\sum_{k_{1},k_{2},k_{3}=1}^{5}\ {\ \Pi }_{k_{1}}{\ \Pi }_{k_{2}}{\ \Pi }%
_{k_{3}}.
\end{equation}

So ${\bf I}_{{\cal D}}$  can be represented by a completely reducible Quiver diagram with $%
d_{1}d_{2}d_{3}=5\times 5\times 5$ vertices. Similar expansion to eqs(4.29)
may be written down for the generators $J_{a}$ of the quantum symmetries.
Setting
\begin{eqnarray}
A_{k_{1}}^{\pm } &=&\ a_{1,k_{1}}^{\pm }\otimes I_{id}\otimes I_{id},
\nonumber \\
A_{k_{2}}^{\pm } &=&\ I_{id}\otimes a_{1,k_{1}}^{\pm }\otimes I_{id}, \\
A_{k_{3}}^{\pm } &=&\ I_{id}\otimes I_{id}\otimes a_{k_{3}}^{\pm },
\nonumber
\end{eqnarray}
we can write for instance $J_{a}$\ and \ $J_{a_{1}}J_{a_{2}}$ as follows;
\begin{equation}
J_{a}=\sum_{k_{a}=1}^{5}\ A_{k_{a}}^{+},\quad
J_{a_{1}}J_{a_{2}}=\sum_{k_{a_{1}}k_{a_{2}}=1}^{5}\
A_{k_{a_{1}}}^{+}A_{k_{a_{2}}}^{+}.
\end{equation}
While the quiver diagram for the $J_{a}$\ are similar to that given by
figures 2.b and c and figures 3, the quiver diagrams associated with\ $%
J_{a_{i}}J_{a_{j}} $\ are obtained by taking cross products and are of
type figure 4.
\begin{figure}[tbh]
\begin{center}
\epsfxsize=14cm \epsffile{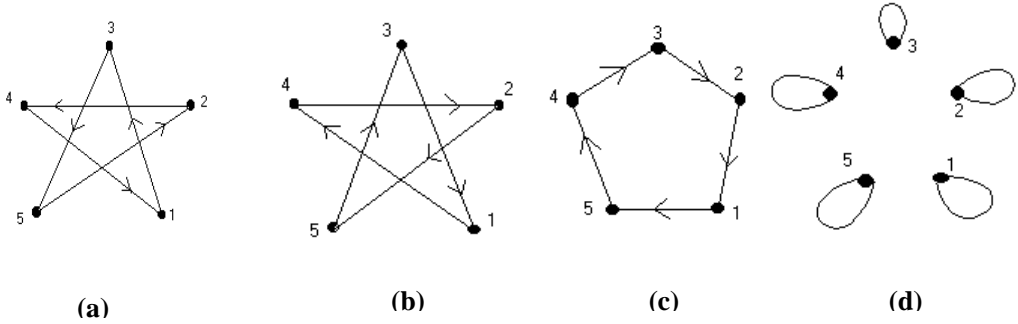}
\end{center}
\caption{{\protect\small {\it Fig 3-a represents the quiver diagram of the $%
Q^{2}$ operator, the oriented links define the independent massless chiral
fields on the $D$ brane at singularity. Fig 3-b represents the quiver
diagram of the $Q^{3}$, which upon reversing the orientation of the links is
similar to fig 3-a. Fig 3-c represents the quiver diagram of the $Q^{4}$
operator, it just the reverse of the diagram of $Q$ and Fig 3-d represents
the completely reducible quiver diagram of the $Q^{5}$. }}}
\end{figure}

\section{CP Algebra}

\qquad The construction we will develop here applies to a wide class
Calabi-Yau orbifolds with torsion and offer a unified description for  the
study of non commutative geometry versus discrete torsion. It explains the
origin of the various possible realizations of non commutative manifolds
encountered in the previous section; see eqs(4.17) and (4.23). The strategy
we will be following in this presentation is: First build the CP algebra for
the quintic by working in the $\Phi _{i}$ frame. We study its centre ${\cal Z%
}\left( {\cal A}\left[ \Phi ,F\right] \right) $ and the structure of the non
commutative points of ${\cal Q}$. Then, we build a class of explicit
solutions for the fields moduli by taking $\Phi _{i}$'s as matrix operators
valued in ${\cal D}\left( {{\bf Z}_{5}^{aut}}^{3}\right) $. Next we
reconsider the $Z$ realization from the point of the CP algebra method and
then build the Morita equivalence between ${\cal A}\left[ Z,F\right] $ and $%
{\cal A}\left[ \Phi ,F\right] $. Finally we consider other frames of the CP
algebra and their Morita equivalence.

\subsection{${\cal Q}\boxtimes {\bf Z}_{5}^{3}$ \ CP Algebra}

\qquad Roughly speaking the CP algebra of the quintic $Q\boxtimes \ {\cal A}%
_{G}$ ,with $G={\bf Z}_{5}^{3},$ is a subspace of the algebra of functionals
${\cal F}\ \left[ \left[ {\cal M}at\left( d_{1}d_{2}d_{3},{\bf C}\right) %
\right] \right] $ on the tensor product of the ${\cal M}at(d_{a},C)$ matrix
spaces$.$ This is a fiber bundle based on $Q$ and whose fiber is given by
the algebra of the representation group ${\cal D}\left( G\right) $ .Elements
of $Q\boxtimes \ {\cal A}_{G}$\ are generated by the matrix fields moduli $%
\Phi _{i}$, $1\leq i\leq 5$, and the ${\cal D}\left( G\right) $ group
generators $F_{a}$, $1\leq a\leq 3$. The $\Phi _{i}$ chiral operators and
the $F_{a}$\ generators satisfy the following algebra; see also eqs(2.18)
and (2.20),

\begin{eqnarray}
\Phi _{i}.\Phi _{j} &=&\Phi _{j}.\Phi _{i},  \nonumber \\
F_{a}.\Phi _{i} &=&\lambda _{ai}\ \Phi _{i}.F_{a}, \\
F_{a}.F_{b} &=&\omega ^{\theta _{ab}}\ F_{b}.F_{a},  \nonumber
\end{eqnarray}
where $\lambda _{ai}$ can be set equal to $\omega ^{q_{i}^{a}}$ as required
by the constraint eq ${F_{a}}^{5}={\bf I}_{{\cal D}\left( G\right) }$. The $%
q_{i}^{a}$'s are the same integers we have been considering so far and, for
simplicity, will be chosen as in eqs(3.4). The $\theta _{ab}$'s are integers
defining the discrete torsion matrix eqs(4.11). They are fixed numbers for
well specified torsions. Actually the relations (5.1) define the $\Phi $
realization of the NC quintic ${\cal Q}^{(nc)}$. A class of solutions for $%
{\cal Q}^{(nc)}$ were already considered in section 3 \ by using the
constrained method. In addition to its direct physical interpretation as
supersymmetric vacuum configurations and it link with $4d$ superpotentials
(2.13), there are other novelties in the $\Phi $ realization are: (1) The
matrix fields moduli $\Phi _{i}$ are commuting operators contrary to the $%
Z_{i}$'s which do not as shown on eqs(3.7). (2) The number of generators in
the CP algebra of the quintic is apparently larger than those appearing in
eqs(3.7). In addition to the $\Phi _{i}$ 's, we have moreover the group
generators as principal actors. In eqs(3.7), the $F_{a}$ group generators
act as discrete outer automorphisms; but in the $\Phi $ realization they are
inner ones.

\qquad To prove that the algebra generated by eqs(5.1) is indeed the non
commutative extension of the quintic eq(3.1), we have to show that: (i) the
centre ${\cal Z(}Q\boxtimes \ {\cal A}_{G}$) of the CP algebra eq(5.1) is
exactly the commutative quintic ${\cal Q}$ given by eq(3.1) and (ii) check
that eqs(5.1) are equivalent to eqs(2.22) and (3.9) used for defining ${\cal Q}^{(nc)}$
\ in the $Z$ realization. Put differently, we have to establish that,

\begin{eqnarray}
Q\boxtimes \ {\cal A}_{G}\ &\simeq &\ {\cal Q}^{(nc)},\qquad \qquad \qquad
\qquad \qquad \qquad \qquad (a)  \nonumber \\
{\cal Z}(Q\boxtimes \ {\cal A}_{G})\ &\simeq &\ {\cal Q}.\qquad \qquad
\qquad \qquad \qquad \qquad \qquad \qquad (b)
\end{eqnarray}

We will first address eq(5.2-a) and then discuss the equivalence.

\subsubsection{ The Centre ${\cal ZA}(Q\boxtimes \ {\cal A}_{G}$)}

\qquad To start note that generic elements $v$ of \ $Q\boxtimes \ {\cal A}%
_{G}$ \ are functions of the matrix generators $\Phi _{i}$ and $F_{a}$; i.e,
\ $v=v(\Phi _{1},\dots ,\Phi _{5};F_{1},F_{2},F_{3})$ . As the ${\bf Z}_{5}$%
' s are cyclic groups, the $v$' s can be usually expanded in a power series
of the ${\cal D}\left( G\right) $ generators as follows;

\begin{equation}
v=\sum_{i,j,k=1}^{5} \ {\cal F}_{ijk}(\Phi_{1},\dots,\Phi_{5})\
F_{1}^{i}.F_{2}^{j}.F_{3}^{k},
\end{equation}

where ${\cal F}_{ijk}(\Phi _{1},\dots ,\Phi _{5})$ are arbitrary analytic
functions of the $\Phi _{l}$ matrix fields. To get the elements $\zeta _{\mu
}=\zeta _{\mu }(\Phi _{1},\dots ,\Phi _{5})$ of the centre \ ${\cal Z}%
(Q\boxtimes \ {\cal A}_{G})$, \ one has to solve the constraint eqs

\begin{equation}
\zeta _{\mu }\ v\ =\ v\ \zeta _{\mu },\ \ \ \forall \ v\in Q\boxtimes \
{\cal A}_{G},
\end{equation}
or equivalently,
\begin{eqnarray}
\zeta _{\mu } \ \Phi _{i} &=&\Phi _{i}\ \zeta _{\mu },  \nonumber \\
\zeta _{\mu }\ F_{a} &=&F_{a}\ \zeta _{\mu }.
\end{eqnarray}

Eq(5.4) shows that, in presence of discrete torsion, \ $\zeta _{\mu }={\cal F%
}_{(0,0,0)}(\Phi _{1},\dots ,\Phi _{5})$\ while eqs(5.5) requires that $%
\zeta _{\mu }$ should be invariant under the ${\bf Z}_{5}^{3}$ actions.
Particular solutions of these constraint eqs, in agreement with the Schur
lemma, are given by,
\begin{eqnarray}
\zeta _{i} &=&\Phi _{i}^{5}=x_{i}^{5}\ {\bf I}_{{\cal D}\left( G\right) }
\nonumber \\
\zeta _{0} &=&\Pi _{i=1}^{5}\ \Phi _{i}=\ {\bf I}_{{\cal D}\left( G\right)
}\ \ (\Pi _{i=1}^{5}\ x_{i}),
\end{eqnarray}

where the $x_{i}$'s are given by (2.3); they are essentially as in eq(3.1).
Therefore the centre ${\cal Z}(Q\boxtimes \ {\cal A}_{G})$ is given by the
linear combination of $\zeta _{\mu }$, namely \ $\sum_{\mu }\zeta _{\mu
}=\sum_{i=1}^{5}(x_{i}^{5}+\Pi _{i=1}^{5}x_{i})\ {\bf I}_{{\cal D}\left(
G\right) }$. \ As such, generic $v$ elements of the quintic CP algebra can
be usually rewritten as:

\begin{equation}
v=\sum_{i=1}^{5}(x_{i}^{5}+\Pi _{i=1}^{5}x_{i}).{\bf I}_{{\cal D}\left(
G\right) }+\sum_{i,j,k=1}^{5}\ {\cal F^{\prime }}_{k_{1}k_{2}k_{3}}(\Phi
_{1},\dots ,\Phi _{5})\ F_{1}^{k_{1}}.F_{2}^{k_{2}}.F_{3}^{k_{3}}.
\end{equation}

Actually this relation shows that non commutative points in the $(Q\boxtimes
\ {\cal A}_{G})$ \ CP algebra of the quintic can be thought of as fibers of
a fiber bundle of base\  $\sum_{i=1}^{5}(x_{i}^{5}+\Pi _{i=1}^{5}x_{i})=0$ and
fiber given by the algebra of the group representation ${\cal D}\left(
G\right) $.

\subsubsection{Structure of \ the Fibration}

\qquad Demanding invariance of the above relation under ${\bf Z}_{5}^{3}$,
we can extract some informations on the structure of the \ ${\cal F^{\prime }%
}_{k_{1}k_{2}k_{3}}$ \ monomials. Indeed, solving the identity $%
F_{a}vF_{a}^{-1}=v$, we get;

\begin{equation}
F_{a}{\cal F^{\prime }}_{k_{1}k_{2}k_{3}}\ F_{a}^{-1}=\ Q_{\left(
k_{1}k_{2}k_{3}\right) }^{a}\ \ {\cal F^{\prime }}_{k_{1}k_{2}k_{3}},
\end{equation}

with
\begin{equation}
\ Q_{\left( k_{1}k_{2}k_{3}\right) }^{a}\ =-\sum_{b=1}^{3}\ k_{b}\ \theta
_{ab}+5{\bf Z}.
\end{equation}

Moreover, using the relation \ $\eta _{ab}l_{ij}=\ q_{i}^{a}q_{j}^{b}$\ that
follows from\ eqs(3.14-15), one can express $\theta _{ab}$ as
\begin{equation}
\theta _{ab}=\sum_{i,j=1}^{5}q_{i}^{a}\left( l_{ij}^{-1}-l_{ji}^{-1}\right)
\ q_{j}^{b}
\end{equation}

Putting back into eq(5.9), we find
\begin{equation}
\ Q_{\left( k_{1}k_{2}k_{3}\right) }^{a}\ =-\sum_{i=1}^{5}q_{i}^{a}\left[
\sum_{j=1}^{5}\sum_{b=1}^{3}\ k_{b}\ \left( l_{ij}^{-1}-l_{ji}^{-1}\right) \
q_{j}^{b}\right] +5{\bf Z.}
\end{equation}

But this relation tells us that the ${\cal F^{\prime }}_{k_{1}k_{2}k_{3}}$ \
monomials have the form

\begin{equation}
{\cal F^{\prime }}_{k_{1}k_{2}k_{3}}=\prod_{i,j=1}^{5}\left( \left[ \left(
\Phi _{i}\right) ^{\left( l_{ij}^{-1}-l_{ji}^{-1}\right) q_{j}^{1}}\right]
^{\ k_{1}}\left[ \left( \Phi _{i}\right) ^{\left(
l_{ij}^{-1}-l_{ji}^{-1}\right) q_{j}^{2}}\right] ^{\ k_{2}}\left[ \left(
\Phi _{i}\right) ^{\left( l_{ij}^{-1}-l_{ji}^{-1}\right) q_{j}^{3}}\right]
^{\ k_{3}}\right)
\end{equation}

Therefore the most general form of the non commutative quintic invariant
under ${\bf Z}_{5}^{3}$ transformation is given by eqs(5.3) with ${\cal %
F^{\prime }}_{k_{1}k_{2}k_{3}}$ as in eqs(5.12).

\subsection{ Explicit Expression of the $\Phi_{i}$'s}

\qquad Explicit solutions for the generators of the $Q\boxtimes \ {\cal A}%
_{G}$\ algebra can be written down by using the commutativity feature of the
$\Phi _{i}$'s. To build a matrix representation, we consider the simple
choice eqs(4.17) for the $F_{a}$ group generators;

\begin{eqnarray}
F_{1}&=& P\otimes Q^{\eta_{12}}\otimes Q^{\eta_{13}},  \nonumber \\
F_{2}&=&Q^{\eta_{21}}\otimes P\otimes Q^{\eta_{23}}, \\
F_{3}&=&Q^{\eta_{31}}\otimes Q^{\eta_{32}}\otimes P ,  \nonumber
\end{eqnarray}

where $P_{1}=P_{2}=P_{3}=P$ and where $P$ and $Q$ satisfy eqs(3.20) with $%
\alpha $ taken as the fifth primitive root of unity. The $F_{1},$ $F_{2}$\
and\ $F_{3}$\ operators are mainly elements of \ ${\cal D}\left( {\bf Z}%
_{5}\otimes {\bf Z}_{5}^{aut}\otimes {\bf Z}_{5}^{aut}\right) $, ${\cal D}%
\left( {\bf Z}_{5}^{aut}\otimes {\bf Z}_{5}\otimes {\bf Z}_{5}^{aut}\right) $
and ${\cal D}\left( {\bf Z}_{5}^{aut}\otimes {\bf Z}_{5}^{aut}\otimes {\bf Z}%
_{5}\right) $ respectively. Using these relations, it is not a difficult to
check that,

\begin{equation}
F_{1}\ \left[ Q^{q_{i}^{1}}\otimes Q^{q_{i}^{2}}\otimes Q^{q_{i}^{3}}\right]
=\omega ^{q_{i}^{1}}\ \left[ Q^{q_{i}^{1}}\otimes Q^{q_{i}^{2}}\otimes
Q^{q_{i}^{3}}\right] \ F_{1}.
\end{equation}

More generally we have the following identities;

\begin{equation}
F_{a}\ \left[ J_{1}^{q_{i}^{1}}.J_{2}^{q_{i}^{2}}.J_{3}^{q_{i}^{3}}\right]
=\omega ^{q_{i}^{a}}\ \left[
J_{1}^{q_{i}^{1}}.J_{2}^{q_{i}^{2}}.J_{3}^{q_{i}^{3}}\right] \ F_{a}.
\end{equation}

Therefore a class of solution for the matrix fields moduli reads as

\begin{eqnarray}
\Phi _{i} &=&z_{i}\ \left[ Q^{q_{i}^{1}}\otimes Q^{q_{i}^{2}}\otimes
Q^{q_{i}^{3}}\right]  \nonumber \\
&=&z_{i}\ \prod_{a=1}^{3}\ J_{a}^{q_{i}^{a}}.
\end{eqnarray}

Such class of solutions belongs to the algebra of the representation of the
\ ${\cal D}\left( {\bf Z}_{5}^{aut}\otimes {\bf Z}_{5}^{aut}\otimes {\bf Z}%
_{5}^{aut}\right) $ representation of the automorphism symmetry. In addition
to the manifest commutativity property $\Phi _{i}.\Phi _{j}=\Phi _{j}.\Phi
_{i}$, it satisfies naturally the relations (5.1-2); due to the property $%
Q^{5}=I_{id}$ \ and to the Calabi-Yau condition eqs(3.5). From these
solutions one also learns that it is possible to build other solutions by
using both \ ${\bf Z}_{5}$\ \ and \ ${\bf Z}_{5}^{aut}$ factors of the full
discrete symmetry $\widehat{G}$. In what follows we will consider these kind
of solutions and show that they are indeed equivalent to eqs(5.13) and
(5.16).

\section{General Solutions}

\qquad Let us summarize what we have been studying so far. The algebraic
relations defining the non commutative quintic ${\cal Q}^{(nc)}$ as appeared
first in $\cite{2}$ read, in the coordinate patch $z_{5}=1$, as;

\begin{eqnarray}
Z_{1}Z_{2} &=&\alpha Z_{2}Z_{1},  \nonumber \\
Z_{1}Z_{3}&=&\alpha ^{-1}\beta Z_{3}Z_{1},  \nonumber \\
Z_{1}Z_{4} &=&\beta^{-1}Z_{4}Z_{1}, \\
Z_{2}Z_{3}&=&\alpha \gamma Z_{3}Z_{2},  \nonumber \\
Z_{2}Z_{4} &=&\gamma ^{-1}Z_{4}Z_{2},  \nonumber \\
Z_{3}Z_{4}&=&\beta \gamma Z_{4}Z_{3},  \nonumber \\
Z_{i}Z_{5} &=&Z_{5}Z_{i},\qquad i =1,2,3,4;  \nonumber
\end{eqnarray}

where $\alpha, \beta, \gamma$ are fifth roots of unity. In $\cite{24}$, it
was noted that the above relations are very special and can generalized as
follows:

\begin{eqnarray}
Z_{i}Z_{j} &=&\beta _{ij}Z_{j}Z_{i};\qquad \ i,j=1,...,(n+1),  \nonumber \\
Z_{i}Z_{d+2} &=&Z_{d+2}Z_{i};\qquad \ i=1,...,(n+1),
\end{eqnarray}

For the case of the quintic the $n$ parameter should be set equal to three; $%
n=3$; but the above relations are also valid for complex $n$-dimension
holomorphic homogeneous Calabi-Yau hypersurfaces. The non commutative
deformation parameters $\beta _{kl}$ are realized in terms of the Calabi-Yau
charges as follows:

\begin{equation}
\beta _{ij}=\exp i\left( \frac{2\pi }{n+2}m_{ab}q_{i}^{a}q_{j}^{b}\right)
=\omega ^{m_{ab}q_{i}^{a}q_{j}^{b}}.
\end{equation}

where $m_{ab}$ is some given matrix with integer entries; it has remained
without interpretation in $\cite{24}$; but here we will identify its origin.
Indeed, using the CP algebra representation of the non commutative quintic,
we will prove that $m_{ab}$ is equal to $\left( \ \eta _{ab}{}^{-1}-\eta
_{ba}{}^{-1}\right) $; \ where $\eta _{ab}$\ is as in eqs(5.13). To that
purpose, let us analyze the conditions under which eqs(5.1) and (6.1-2) are
equivalent.

\subsection{Morphisms between Eqs(5.1) and Eqs(6.1)}

\qquad At first sight, saying that eqs (5.1) and eqs(6.1-2) have to
equivalent seems a little bit ambiguous as the naive counting of the degrees of freedom seems not maching. Eqs (6.1-2) associated with the
constrained formulation of the non commutative quintic, involve only the
space variable operators $Z_{i}$; the $F_{a}$ group generators do not appear
manifestly; they act as outer automorphisms on eqs (6.1-2). This is however
not a major problem since it can\ be overcome if one supposes that in the
constrained form, the $Z_{i}$'s and the $F_{a}$'s do not couple each others
and so eqs(6.2) should be rather understood as:

\begin{eqnarray}
Z_{i}Z_{j} &=&\beta_{ij} \ Z_{j}Z_{i};\qquad \ i,j=1,...,(d+1),  \nonumber \\
Z_{i}Z_{d+2}&=&Z_{d+2}Z_{i};\qquad \ i=1,...,(d+1),
\end{eqnarray}

together with
\begin{equation}
F_a.F_b=\omega^{\theta_{ab}} \ F_{b}.F_a ,
\end{equation}
and moreover,

\begin{equation}
F_{a}.Z_{i}=Z_{i}.F_{a}.
\end{equation}

Though trivial, eqs(6.6) are however very important for us; they will
determine the morphism linking eqs(5.1) and eqs(6.1-2). To do so, let us
first introduce, $\Psi _{L}$ and $\Psi _{R}$ , two maps describing
respectively left and right morphisms of the algebra eqs(5.1):

\begin{eqnarray}
\Psi_{L}:\ \Phi_{i}\ &\longrightarrow &\ Z_{i}=\Gamma_{i}.\Phi_i, \\
\Psi_{R}:\ \Phi_{i}\ &\longrightarrow &\ Z^{\prime}_{i}= \Phi_i.\Gamma_{i},
\end{eqnarray}

where $\Gamma _{i}$ is an invertible operator depending on the generators of
the algebra; $\Gamma _{i}=\Gamma _{i}(\Phi _{1},\ldots ,\Phi _{5};F_{a})$.
Here below, we will take the set of $\ \Gamma _{i}^{\prime }$ s in the group
representation ${\cal D}\left( G\right) $ as,

\begin{eqnarray}
\Gamma _{i} &=&\ [P^{p_{i}^{1}}\otimes P^{p_{i}^{2}}\otimes P^{p_{i}^{3}}];
\nonumber \\
&=&\ E_{1}^{p_{i}^{1}}.E_{2}^{p_{i}^{2}}.E_{3}^{p_{i}^{3}},
\end{eqnarray}

where the $p_{i}^{a}$\ 's are for the moment some given integers modulo
five; they will be determined by requiring the relations eqs(6.4) to be
hold. Now putting eqs(6.9) back into eqs(6.7-8) and using the realization of
$\Phi _{i}$ given by eq(5.16) as well as the eqs(5.14-15), one gets amongst
others;

\begin{equation}
Z^{\prime}_{i}=\omega^{-\sum_{a=1}^{3} {p_{i}^{a}}{q_{i}^{a}}}\ Z_{i}.
\end{equation}

Actually this relation shows that $\Psi _{L}$\ and $\Psi _{R}$\ are related;
$\Psi _{R}=(\omega ^{-\sum_{a=1}^{3}{p_{i}^{a}}{q_{i}^{a}}})\Psi _{L}$;\ \
so we will focus our attention here after  on the left morphism $\Psi _{L}$ \
only.

Next, we solve $F_{a}.Z_{i}=Z_{i}.F_{a}$ by using eqs(5.13) and eqs(6.7). We
find:

\begin{equation}
\sum_{b=1}^{3}\eta_{ab}.p^{b}_{i}=q^{a}_{i},
\end{equation}

where $\eta _{ab}$ is the same matrix we have been using. This relation
shows that the $p_{i}^{a}$ are related to the $q_{i}^{a}$ via the torsion
matrix namely $p_{i}^{a}=\sum_{b=1}^{3}\eta _{ab}^{-1}q_{i}^{b}$. \ Since
the $p_{i}^{a}$\ are integers, eq(6.11) requires that the matrix $\ \eta
_{ab}$ $\ $to belong to\ $\ Sl(3,{\bf Z})$ and shows moreover that $%
p_{i}^{a} $ ' s satisfy themselves the Calabi-Yau condition:

\begin{equation}
\sum_{1=1}^{5}\ p^{a}_{i}=0.
\end{equation}

Furthermore eqs(6.11) show that, whenever torsions are present, the
commutative quintic admits an extra hidden discrete symmetry acting on the $%
z_{i}$'s as

\begin{equation}
z_{i}\ \ \to \ \ \omega^{p^{a}_{i}}\ z_{i}.
\end{equation}

Therefore the $\Psi _{L}$ morphism is torsion dependent and is given by the
left multiplication operators;

\begin{equation}
\Gamma _{i}=\prod_{a=1}^{3}\ (P{\bf _{a}}^{-\sum_{b=1}^{3}\eta _{ab}^{-1}\
q_{i}^{b}}).
\end{equation}

This is a completely reducible operator suggesting that the $\Psi _{L}$
morphism can be also written as $\Psi _{L}=\Psi _{1}\otimes \Psi _{2}\otimes \Psi
_{3}$, with $\Psi _{a}=\oplus _{i=1}^{5}\Psi _{a,i}$ and $\Psi _{a,s}={%
\alpha }^{-s\sum_{b=1}^{3}\eta _{ab}^{-1}\ q_{i}^{b}}$. \ The $\Psi _{L}$
morphism depends on the Calabi-Yau charges $q_{i}^{a}$ and the inverse of
the discrete torsion matrix $\eta _{ab}^{-1}$. \ Using these expressions of
and combining altogether the above relations, we get the $Z_{i}$ realization
for the CP algebra of the non commutative quintic, namely
\begin{eqnarray}
Z_{i} &=&P^{p_{i}^{1}}Q^{q_{i}^{1}}\otimes P^{p_{i}^{2}}Q^{q_{i}^{2}}\otimes
P^{p_{i}^{3}}Q^{q_{i}^{2}},  \nonumber \\
&=&\prod_{a=1}^{3}E_{a}^{p_{i}^{a}}J_{a}^{q_{i}^{a}},
\end{eqnarray}

where $p_{i}^{a}$ are as in eqs(6.11-12). Finally we calculate the relation
between $Z_{i}Z_{j}$ and $Z_{j}Z_{i}$. Using eqs(5.16) and eqs(6.7-9), we
show, by straightforward computations, that

\begin{equation}
Z_{i}Z_{j}= \omega^{\sum_{a=1}^{3}\ [p^{a}_{i}q^{a}_{j}-p^{a}_{j}q^{a}_{i}]}
\ Z_{j}Z_{i}
\end{equation}

Comparing this relation with eqs(6.3), one gets $%
m_{ab}\ q_{i}^{a}q_{j}^{b}=[p_{i}^{a}q_{j}^{a}-p_{j}^{a}q_{i}^{a}]$, then
using the inverse eq(6.11) by expressing $p_{i}^{a}$ in terms of the $%
q_{i}^{a}$, we can rewrite the previous relation as $%
m_{ab}q_{i}^{a}q_{j}^{b}=\left( \eta _{ba}^{-1}-\eta _{ab}^{-1}\right)
q_{i}^{a}q_{j}^{b}$\ showing that $m_{ab}=-\left( \eta _{ab}^{-1}-\eta
_{ba}^{-1}\right) $. In what follows we collect the various relations one gets,

\begin{eqnarray}
l_{ij} &=&\ \sum_{a=1}^{3}\ p_{i}^{a}q_{j}^{a}=\sum_{a,b=1}^{3}\eta
_{ab}^{-1}q_{i}^{b}q_{j}^{a}=\sum_{a=1}^{3}\ \eta _{ab}p_{i}^{a}p_{j}^{b}
\nonumber \\
L_{ij} &=&\ l_{ij}-l_{ji}=p_{i}^{a}q_{j}^{a}-p_{j}^{a}q_{i}^{a},  \nonumber
\\
L_{ij} &=&\ (\eta _{ab}-\eta _{ba})\ p_{i}^{a}p_{j}^{b}, \\
L_{ij} &=&\ -(\eta _{ab}^{-1}-\eta _{ba}^{-1})\ q_{i}^{a}q_{j}^{b}.
\nonumber
\end{eqnarray}
We end this discussion on the non commutative quintic representations by
noting that given: (i) a set of Calabi-Yau charges $q_{i}^{a}$, defining the
charges of $\ {{\bf Z}_{5}^{3}}$, and \ (ii) an invertible matrix $\eta
_{ab} $ of \ $SL(3;Z)$ \ with a non zero antisymmetric part; we can build
various, but equivalent, realizations of the CP algebra of the NC quintic.
Starting from eqs(5.16), perform variables change type;

\begin{eqnarray}
\Phi _{i}^{\prime } &=&\Omega _{i}\text{ }\Phi _{i},  \nonumber \\
\Phi _{i}^{\prime } &=&\tau _{i}\text{ }\Phi _{i}\text{ }\Omega _{i},
\nonumber \\
\Omega _{i}\Omega _{j} &=&\varepsilon _{ij}\ \Omega _{j}\text{ }\Omega _{i},
\\
F_{a}\Omega _{i} &=&\kappa _{ai}^{-1}\ \Omega _{i}\text{ }F_{a},  \nonumber
\end{eqnarray}

where $\Omega _{i}=\Omega _{i}(E_{a},J_{a})$ are given monomials of \ $E_{a}$
and $\ J_{a}$ with structure constants $\tau _{i},\varepsilon _{ij}$ and $%
\kappa _{ai}$. Then compute the commutation relations for the new operators,
we get;

\begin{eqnarray}
\Phi _{i}^{\prime }\Phi _{j}^{\prime } &=&\tau _{i}^{-1}\varepsilon
_{ij}\tau _{j}\ \text{\ }\Phi _{j}^{\prime }\Phi _{i}^{\prime },  \nonumber
\\
F_{a}\Phi _{i}^{\prime } &=&\kappa _{ai}^{-1}\lambda _{ai}\ \text{\ }\Phi
_{i}^{\prime }F_{a}, \\
F_{a}F_{b} &=&t_{ab}\text{ \ }F_{b}F_{a},  \nonumber
\end{eqnarray}
where the \ new generators $\Phi _{i}^{\prime }$ \ do no longer commute
among themselves nor with the group representation generators. \ These
relations are realizations of the NC quintic interpolating between the \ $%
\Phi _{i}$ realization and the $Z_{i}$ one.

\subsection{Example}

\qquad To fix the ideas, let consider an example by\ choosing $\ $\ $\Omega
_{i}$\ \ as\ $\ F_{1}^{r_{1}^{a}}F_{2}^{r_{2}^{a}}F_{3}^{r_{3}^{a}}$ \ with $%
r_{i}^{a}$ some integers. In this case,\ the above mentioned structure
constants read as;
\begin{eqnarray}
\tau _{i} &=&\ \omega ^{\sum_{a}{r_{i}^{a}q_{i}^{a}}},  \nonumber \\
\varepsilon _{ij} &=&\ \omega ^{\theta _{ab\text{ }}r_{i}^{a}\text{ }%
r_{j}^{b}}, \\
\kappa _{ai} &=&\ \omega ^{-\theta _{ab}\ {r_{i}^{b}}},  \nonumber
\end{eqnarray}

As we see, the structure constants $\varepsilon _{ij}$ and $\kappa _{ai}$
are torsion dependent; if $\ \theta _{ab}=0$; \ then $\varepsilon _{ij}=1$ \
and $\kappa _{ai}=1$. Note that in the $\{\Phi _{i}^{\prime };F_{a}\}$
frame, the generators of the CP algebra of the quintic do not commute in
general; except for the following special cases where the non commutative
quintic takes remarkable forms:

{\it (1)}{{\bf \ $\Phi _{i}$ }}{\it Frame: eqs(5.1) and eqs(5.16)}

This is recovered by setting $\tau _{i}^{-1}\varepsilon _{ij}\tau _{j}=1$;
that is,
\begin{equation}
\varepsilon _{ij}=\tau _{i}\tau _{j}^{-1},\quad \kappa _{ai}=1.
\end{equation}

From eq(6.14), one sees that $\Omega _{i}$\ is just the inverse of $\Gamma
_{i}$\ ; that is, $\Omega _{i}=\Gamma _{i}^{-1}.$

{(2) {\bf $Z_{i}$ }}{\it Frame}:{\it \ eqs(6.1-2) and eqs(6.15)}

It is obtained by requiring the {{\bf $Z_{i}$}}\ and $F_{a}$\ operators to
commute; this is equivalent to setting\  $\kappa _{ai}\sigma _{ai}=1$;

\begin{eqnarray}
\sigma _{ai} &=&\kappa _{ai}^{-1},\quad \beta _{ij}=\tau
_{i}^{-1}\varepsilon _{ij}\tau _{j},  \nonumber \\
\Omega _{i} &=&\Gamma _{i}.
\end{eqnarray}
Since the two sets of matrix generators \ $\{\Phi _{i}^{\prime }=Z_{i}\}$ \
and \ $\{F_{a}\}$ decouple completely, the NC quintic is then described by a
trivial fibration of the group algebra ${\cal A}_{G}$\ \ on the commutative
quintic as shown here below;

\begin{equation}
Q\boxtimes \ {\cal A}_{G}\equiv {\cal A}[\Phi _{i}\text{ };\text{ }%
F_{a}]\equiv {\cal A}[Z_{i}]\otimes {\cal A}[F_{a}]\equiv {\cal Q}%
^{nc}\otimes {\cal A}_{G}.
\end{equation}

This relation shows that the constrained method we studied in section 3 and
the CP algebra approach we have been considering in sections 5 and 6 are
Morita equivalent.

\section{ \ \ Fractional Branes}

\qquad The realization of the NC quintic we have studied here above concerns
only the regular points of the algebra, that is non singular ones. In this
section, we want to complete this analysis by considering the
representations for singular points. This is not only important for the study of fractional branes at singularities\ but also for  answering the question regarding the nature of fractional branes on the NC quintic and more generally on NC
Calabi-Yau hypersurfaces.

\qquad To do so, we shall first determine and classify the various sets $%
{\cal S}_{(\mu )}$ of singular points of orbifolds of the quintic; then we
give the corresponding singular solutions. At first sight and as far as the
full ${\bf Z}_{5}^{3}$  geometric symmetry of ${\cal Q}$  is concerned, we have only one fixed point under the
${\bf Z}_{5}^{3}$ actions, namely $(z_{1},z_{2},z_{3},z_{4},1)=(0,0,0,0,1).$
This point belongs however to the ${\bf CP}^{4}$ projective space; but does
not belong to the quintic ${\cal Q}$; no point of the quintic is then fixed
by the full symmetry. This property is valid for all Calabi-Yau
hypersurfaces; no point of complex $n$ dimensional Calabi-Yau hypersurfaces
\ ${\cal P}{(z_{1},\dots ,z_{n+2})}$\ is fixed under the full ${\bf Z}%
_{n+2}^{n}$ invariance. We will therefore consider points of ${\cal Q}$ that
are fixed under subgroups $G_{\left[ \alpha \right] }$ of $\ {\bf Z}_{5}^{3}$%
. Then we describe the various fractional branes living at these
singularities, the corresponding quiver diagrams and the massless chiral
fields of the effective theory on the D branes.

\qquad As there are several subgroups $G_{\left[ \alpha \right] }$ in $\
{\bf Z}_{5}^{3}$, we shall fix our attention on two categories of
subsymmetries; those isomorphic to\ ${\bf Z}_{5};$\ i.e \ $G_{\left[ 1\right]
}\simeq {\bf Z}_{5}$ and those isomorphic to ${\bf Z}_{5}^{2};$\ i.e \ $G_{%
\left[ 2\right] }\simeq {\bf Z}_{5}^{2}$. The Calabi-Yau charges will be
taken as in eqs(3.5). Generalization to subgroups of ${\bf Z}_{n+2}^{n}$,
though tedious, is a priori straightforward.

\subsection{Fixed subspaces of ${\cal Q}^{\left[ \protect\alpha \right] }$}

\qquad We will consider first the spaces ${\cal S}_{(a)}$ of \ fixed points
under a generic ${\bf Z}_{5}$ factor of ${\bf Z}_{5}^{3}$. Then we examine
the spaces ${\cal S}_{(ab)}$ of fixed points under$\ {\bf Z}_{5}^{2}$ \
factors. To have an idea on what these spaces look like, it is interesting
to think about the quintic homogeneous hypersurface eq(3.1) as a fiber
bundle described by the following equation;

\begin{equation}
P(z_{1},\ldots ,z_{5})=\sum_{n_{m+1}...n_{m+5}=0}^{5}\
b_{n_{m+1}...n_{m+5}}\ z_{i_{m+1}}^{n_{m+1}}\dots z_{i_{m+5}}^{n_{m+5}},
\end{equation}
where $b_{n_{m+1}...n_{m+5}}=b_{n_{m+1}...n_{m+5}}(z_{i_{1}},\ldots
,z_{i_{m}})$ are some given monomials, in the $z_{i_{1}},\ldots ,z_{i_{m}}$
\ complex variables, with appropriate degrees. Let us give examples on how
this works in practice.

\subsubsection{${\bf CP}^{2}\bowtie {\cal S}_{1}${\it \ }Fibration}

\qquad A simple example of realizing fibrations of the quintic consists to
rewrite eq(7.1) as,

\begin{equation}
P(z_{1},\ldots ,z_{5})=b_{00}+b_{11}\text{ }z_{1}z_{2}+b_{50}\text{ }%
z_{1}^{5}+b_{05}\text{ }z_{2}^{5},
\end{equation}
where the $b_{mn}$\ coefficient functions are given by
\begin{eqnarray}
b_{00}(z_{3},z_{4},z_{5}) &=&z_{3}^{5}+z_{4}^{5}+z_{5}^{5},  \nonumber \\
b_{11}(z_{3},z_{4},z_{5}) &=&\ a_{0}z_{3}.z_{4}.z_{5},  \nonumber \\
b_{50}(z_{3},z_{4},z_{5}) &=&1, \\
b_{05}(z_{3},z_{4},z_{5}) &=&1,  \nonumber
\end{eqnarray}
and all remaining others are equal to zero. Eqs(7.2-3) mean that the quintic
may be viewed as a fibration space whose base ${\bf B_{1}}$ is just the $%
{\bf CP}^{2}$ space and whose complex one dimension fiber ${\bf F}$, to
which we refer here below to as ${\cal S}_{1}$, is given by,

\begin{equation}
z_{1}^{5}+z_{2}^{5}+b_{1}\ z_{1}z_{2}=0.
\end{equation}

This relation is invariant under the change \ $\left( \ z_{1},\text{ }%
z_{2}\right) $ $\ \rightarrow $ $\ \ \left( \ \omega \text{ }z_{1},\text{ }%
\omega ^{-1}z_{2}\right) ;$ that is under the ${\bf Z}_{5}$\ subsymmetry of
charges $q_{i}^{1}$; the \ ${\bf B}_{1}$ base space is not viewed at all
under this change. The symmetry of the\ ${\cal S}_{1}$\ fiber has one fixed
point namely $\left( 0,0\right) $ and so \ ${\cal S}_{1}$ is singular at the
origin $\ z_{1}=z_{2}=0$. \ To see what eq(7.4) represents, note that from
the $z_{1}$\ and $z_{2}$\ variables, one can build three invariant namely $%
u=z_{1}^{5}$, $v=z_{2}^{5}$ and $\ w=z_{1}z_{2}$ having an ${\bf A}_{4}$\
singularity. In terms of the new variables, the equation of the ${\cal S}%
_{1} $\ complex curve reads as
\begin{eqnarray}
u+v+b\text{ }w &=&0  \nonumber \\
uv &=&w^{5}.
\end{eqnarray}

Therefore near the fixed point $z_{1}=z_{2}=0$, the ${\bf Z}_{5}$\  orbifold of the  commutative quintic $%
{\cal Q}$ can be then viewed as given by the fiber bundle ${\bf CP}%
^{2}\bowtie {\cal S}_{1}$ with a vanishing two cycle at $z_{1}=z_{2}=0$.
Before going ahead, let us comment briefly the complex resolution of this
kind of singularity and give its toric geometry diagram representation as
shown here below.

\begin{equation}
\begin{tabular}{|l|l|}
\hline
${\bf A}_{2}$ Singularity & $xy=z^{2}$ \\ \hline
${\bf A}_{4}$ Singularity & $uv=w^{5}$ \\ \hline
\begin{tabular}{l}
Complex \\
Resolution of ${\bf A}_{4}$%
\end{tabular}
& $uv=w^{5}+\alpha _{4}w^{4}+\alpha _{3}w^{3}+\alpha _{2}w^{2}+\alpha
_{1}w+\alpha _{0}$ \\ \hline
Rules &
\begin{tabular}{l}
$i)$ \ \ White\ nodes such as\ $\ \bigcirc $ \ , are associated \\
\ \ \ \ \ to each non compact $\ {\bf C}$ variables $x$ and $y$ \\
$ii)$ \ \ Nodes such as \ $\otimes $ \ are associated with \\
\ \ \ \ \ \ blown up spheres with self intersection $\left( -2\right) $ \\
$iii)$ \ \ Each link\ \ \ $\longleftrightarrow $\ \ represents intersecting
\\
\ \ \ \ \ \ \ spheres with a weight $\left( 1\right) $%
\end{tabular}
\\ \hline
Quiver Diagrams &
\begin{tabular}{l}
$i)$ \ Quiver diagram for the resolution of ${\bf A}_{2}$: \\
\ \ \ \ \ \ \ \ \ \ \ \ \ \ $\bigcirc _{z^{2}}\longrightarrow \otimes
_{z}\longleftarrow \bigcirc _{1}$ \\
$ii)$ \ Quiver diagram for the resolution of ${\bf A}_{4}$ \\
$\bigcirc _{w^{5}}\longleftrightarrow \otimes _{w^{4}}\longleftrightarrow
\otimes _{w^{3}}\longleftrightarrow \otimes _{w^{2}}\longleftrightarrow
\otimes _{w}\longleftarrow \bigcirc $ $_{1}$%
\end{tabular}
\\ \hline
\end{tabular}
\end{equation}

More details on this diagram representation of the Kahler and complex
resolution of $ADE$ singularities as well as applications in string
compactifications may be found in [42-45]

\subsubsection{${\bf B}\bowtie {\cal S}_{12}${\it \ \ }Fibration}

\qquad The other example we want to give corresponds to the fibration ${\bf B%
}\bowtie {\bf F}\equiv {{\bf CP}^{1}\bowtie }{\cal S}_{12}$. In this case,
the analogue of the above equations read for ${\cal S}_{12}$ as follows:

\begin{equation}
P(z_{1},\dots ,z_{5})=b_{000}+b_{111}\text{ }z_{1}z_{2}z_{3}+b_{500}\text{ }%
z_{1}^{5}+b_{050}\text{ }z_{2}^{5}+b_{005}\text{ }z_{3}^{5},
\end{equation}
where the $b_{mnr}$\ coefficients are follows,
\begin{eqnarray}
b_{000}(z_{4},z_{5}) &=&z_{4}^{5}+z_{5}^{5},  \nonumber \\
b_{111}(z_{4},z_{5}) &=&\ a_{0}z_{4}.z_{5},  \nonumber \\
b_{500}(z_{4},z_{5}) &=&1, \\
b_{050}(z_{4},z_{5}) &=&1,  \nonumber \\
b_{005}(z_{4},z_{5}) &=&1,  \nonumber
\end{eqnarray}
and all others are equal to zero. The equation of the complex two
dimensional singular fiber ${\cal S}_{12}$ is;

\begin{equation}
z_{1}^{5}+z_{2}^{5}+z_{3}^{5}+b\ z_{1}z_{2}z_{3}=0.
\end{equation}

This is a complex two dimension surface invariant under the change \
\begin{equation}
\left( \ z_{1,}\text{ }z_{2},\text{ }z_{3}\right) \ \rightarrow \ \ \left( \
\omega ^{2}\text{ }z_{1,}\text{ }\omega ^{-1}z_{2},\text{ }\omega
^{-1}z_{3}\right) ;
\end{equation}
that is under the ${\bf Z}_{5}\otimes {\bf Z}_{5}$\ subsymmetry of charges $%
\left( q_{i}^{1},q_{i}^{2}\right) $. The latter does not affect the \ ${\bf B%
}_{12}$ base space but has one fixed fiber point at $\left( 0,0,0\right) $;
so \ ${\cal S}_{12}$ is singular at the origin $\ z_{1}=z_{2}=z_{3}=0$. Now
introducing the following four invariant $u_{1}=z_{1}^{5}$, $u_{2}=z_{2}^{5}
$, $u_{3}=z_{3}^{5}$ and $t=z_{1}z_{2}z_{3}$, one sees that they are related
as
\begin{equation}
u_{1}u_{2}u_{3}=t^{5};
\end{equation}

while the ${\cal S}_{12}$ complex surface reads in terms of these invariants
as
\[
u_{1}+u_{2}+u_{3}+bt=0.
\]

From this relation, one recognizes the two individual singularities
associated with each factor of the ${\bf Z}_{5}\otimes {\bf Z}_{5}$
symmetry. These are given by the following ${\bf A}_{4}$\ eqs;
\begin{eqnarray}
u_{1}u_{2} &=&\frac{t^{5}}{u_{3}},\quad \text{for \ }u_{3}\neq 0,  \nonumber
\\
u_{1}u_{3} &=&\frac{t^{5}}{u_{2}},\quad \text{for \ }u_{2}\neq 0.
\end{eqnarray}

Eq(7.11) describes the case where both of the above singularities collapse;
it has a nice description in terms of quivers diagrams.

\bigskip
\begin{equation}
\begin{tabular}{|l|l|}
\hline
Singularity & $u_{1}u_{2}u_{3}=t^{5}$ \\ \hline
\begin{tabular}{l}
Complex \\
Resolution
\end{tabular}
& $u_{1}u_{2}u_{3}=t^{5}+\alpha _{4}t^{4}+\alpha _{3}t^{3}+\alpha
_{2}t^{2}+\alpha _{1}t+\alpha _{0}$ \\ \hline
Rules & Same rules as in previous table. \\ \hline
Symmetries &
\begin{tabular}{l}
\\
${\bf Z}_{5}:\qquad \ \ z_{1}\rightarrow \omega z_{1}$, $z_{3}\rightarrow
\omega ^{-1}z_{3}$,$\quad \quad z_{1}z_{3}$ \ is an invariant. \\
${\bf Z}_{5}\otimes {\bf Z}_{5}:z_{1}\rightarrow \omega ^{2}z_{1}$, $%
z_{2}\rightarrow \omega ^{-1}z_{2}$, $z_{3}\rightarrow \omega ^{-1}z_{3}$,
\\
${\bf Z}_{5}:\qquad \ \ z_{1}\rightarrow \omega z_{1}$, $z_{2}\rightarrow
\omega ^{-1}z_{2}$,\quad $\quad z_{1}z_{2}$ \ is an invariant.
\end{tabular}
\\ \hline
\begin{tabular}{l}
Quiver \\
Diagram
\end{tabular}
&
\begin{tabular}{l}
\\
$\ \ \ \ \ _{\swarrow }\otimes _{\frac{t^{5}}{z_{1}z_{2}}%
}\longleftrightarrow \otimes _{\frac{t^{5}}{\left( z_{1}z_{2}\right) ^{2}}%
}\longleftrightarrow \otimes _{\frac{t^{5}}{\left( z_{1}z_{2}\right) ^{3}}%
}\longleftrightarrow \otimes _{\frac{t^{5}}{\left( z_{1}z_{2}\right) ^{4}}%
}\leftrightarrow \bigcirc $ $_{u_{3}}$:${\bf Z}_{5}$ \\
$\bigcirc _{t^{5}}\longleftrightarrow \otimes _{t^{4}}\longleftrightarrow
\otimes _{t^{3}}\longleftrightarrow \otimes _{t^{2}}\longleftrightarrow
\otimes _{t}\leftrightarrow \bigcirc _{1}$:$\qquad \qquad {\bf Z}_{5}^{2}$
\\
\ \ \ \ \ $^{\nwarrow }\otimes _{\frac{t^{5}}{z_{1}z_{3}}%
}\longleftrightarrow \otimes _{\frac{t^{5}}{\left( z_{1}z_{3}\right) ^{2}}%
}\longleftrightarrow \otimes _{\frac{t^{5}}{\left( z_{1}z_{3}\right) ^{3}}%
}\longleftrightarrow \otimes _{\frac{t^{5}}{\left( z_{1}z_{3}\right) ^{4}}%
}\leftrightarrow \bigcirc $ $_{u_{2}}$:${\bf Z}_{5}$ \\
\\
Here it is represented the three graphs associated to the resolution \\
of the singularities of the discrete symmetries reported above.
\end{tabular}
\\ \hline
\end{tabular}
\end{equation}

Note that the ${\bf Z}_{5}\otimes {\bf Z}_{5}$\ symmetry has a total charge
charge ${\bf q}^{1}+{\bf q}^{2}=\left( 2,-1,-1,0,0\right) $, behaving then
as the ${\bf Z}_{5}$ diagonal symmetry; the remaining off diagonal factor
has a total charge ${\bf q}^{1}-{\bf q}^{2}=\left( 0,-1,1,0,0\right) $.

\subsubsection{Other Fibrations}

\qquad Following the same method we have used for ${\cal S}_{1}$ and ${\cal S%
}_{12}$, one can work out the other ${\bf B}_{a}\bowtie $\ ${\cal S}_{a}$
and \ ${\bf B}_{\left( ab\right) }\bowtie $\ ${\cal S}_{\left( ab\right) }$\
\ Quintic fibrations associated with the natural subgroups of ${\bf Z}%
_{5}^{3}$. Denoting the various invariants under  the subgroups of $%
{\bf Z}_{5}^{3}$ as $u_{i}=z_{i}^{5}$, $w_{ij}=z_{i}z_{j}$ and $\
t_{ijk}=z_{i}z_{j}z_{k}$, one can work out the different equations of the $%
{\cal S}_{a}$ and fibers ${\cal S}_{\left( ab\right) }$; the basis ${\bf B}%
_{a}$\ and $\ {\bf B}_{\left( ab\right) }$ \ are respectively given by the $%
{\bf CP}^{2}$ and ${\bf CP}^{1}$ complex projective spaces. The results are
collected in the table eq(7.14).

\bigskip

\begin{equation}
\begin{tabular}{|l|l||l|l|}
\hline
Fibers & Equations & Fibers & Equations \\ \hline
${\cal S}_{1}$ &
\begin{tabular}{l}
$u_{1}+u_{2}+b_{1}\text{ }w_{12}=0$ \\
$u_{1}u_{2}=w_{12}^{5}.$%
\end{tabular}
\qquad & ${\cal S}_{12}$ &
\begin{tabular}{l}
$u_{1}+u_{2}+u_{3}+b$ $t_{123}=0$, \\
$u_{1}u_{2}u_{3}=t_{123}^{5}$.
\end{tabular}
\\ \hline
${\cal S}_{2}$ &
\begin{tabular}{l}
$u_{1}+u_{3}+b_{2}\text{ }w_{13}=0$ \\
$u_{1}u_{3}=w_{13}^{5}.$%
\end{tabular}
\qquad & ${\cal S}_{23}$ &
\begin{tabular}{l}
$u_{1}+u_{3}+u_{4}+bt_{134}=0$, \\
$u_{1}u_{3}u_{4}=t_{134}^{5}$.
\end{tabular}
\\ \hline
${\cal S}_{3}$ &
\begin{tabular}{l}
$u_{1}+u_{4}+b_{3}\text{ }w_{14}=0$ \\
$u_{1}u_{4}=w_{14}^{5}.$%
\end{tabular}
\qquad & ${\cal S}_{13}$ &
\begin{tabular}{l}
$u_{1}+u_{2}+u_{4}+bt_{124}=0$, \\
$u_{1}u_{2}u_{4}=t_{124}^{5}$.
\end{tabular}
\\ \hline
\end{tabular}
\end{equation}
\bigskip

\bigskip

Having these results at hand, we turn now to give some details by studying
the fixed spaces under the orbifold subgroups ${\bf Z}_{5}$\ and ${\bf Z}%
_{5}^{2}$. We first consider the orbifolds ${\cal Q}^{\left[ 1\right]
}\simeq {\cal R^{\prime }}/{\bf Z}_{5}$ and then the \ ${\bf Z}_{5}^{2}$
orbifolds ${\cal Q}^{\left[ 2\right] }\simeq {\cal R^{\prime \prime }}/{\bf Z%
}_{5}^{2}.$

These orbifolds corresponds also to start from eq(3.1) and choose either
 one of the three $q_{i}^{a}$ vector charges  non vanishing say $%
q_{i}^{1}=\left( 1,-1,0,0,0\right) $ while the two others $%
q_{i}^{2}=q_{i}^{3}={\bf 0}$;\  or two vector charges non vanishing
while the third is zero such as, for instance, \ $q_{i}^{1}=\left(
1,-1,0,0,0\right) ,$\ \ $q_{i}^{2}=\left( 1,0,-1,0,0\right) $\ and\ $%
q_{i}^{3}={\bf 0}$.

\subsection{ Fractional Branes on ${\cal Q}^{\left[ 1\right] }$}

\qquad As there are three manifest ${\bf Z}_{5}$ subsymmetry factors in the
orbifold group ${\bf Z}_{5}^{3}$, each one generated by an operator $F_{a}$,
one can write down three corresponding fibrations for the commutative
quintic, to which we shall refer hereafter to as ${\bf B}_{a}\bowtie $\ $%
{\cal S}_{a}$. \ The ${\bf B}_{a}$ spaces are the bases of the fibration and
$\ $the ${\cal S}_{a}$'s are the fibers encountered above.

\subsubsection{Example : ${\cal Q}^{\left[ 1\right] }\simeq {\bf B}%
_{1}\bowtie $\ ${\cal S}_{1}$}

\qquad Consider the ${\bf Z}_{5}$\ \ subgroup generated by $E_{1}$ with $%
q_{i}^{1}$ charges taken as;
\begin{equation}
q_{i}^{1}=\left( 1,-1,0,0,0\right) ,
\end{equation}
the ${\bf B}_{1}\bowtie $\ ${\cal S}_{1}$ \ fibration of the quintic is just
that given by eqs(7.2-4). Since we are working in the coordinate patch $%
z_{5}=1$, the ${\bf B}_{1}$\ base is a patch of ${\bf CP}^{2}$; that is $\
{\bf B}_{1}\sim {\bf C}^{2}$ parameterized by the $z_{3}$ and $z_{4}$ \
complex coordinates. The codimension two fiber ${\cal S}_{1}$ is given by
the following complex\ curve with an ${\bf A}_{4}$ singularity,
\begin{equation}
u+v+\frac{v^{5}}{u}=0
\end{equation}

\subsubsection{First Results}

\qquad The ${\bf B}_{a}\bowtie $\ ${\cal S}_{a}$\ fibrations of the quintic
are completely determined by the $q_{i}^{a}$ Calabi-Yau charges; the\ ${\bf B%
}_{a}$\ base manifolds are parameterized by those holomorphic coordinates $%
z_{i}$ with $q_{i}^{a}=0$ while the ${\cal S}_{a}$ fibers are parameterized
by those complex variables with non zero $q_{i}^{a}$ charges with fixed
points at the origin. For the $\ q_{i}^{a}$ 's taken as in eq(3.5), we have
the following results collected in the table eq(7.17).

\bigskip

\begin{equation}
\begin{tabular}{|l|l|l|l|}
\hline
Generators $F_{a}$ & $\ \ \ \ \ \ \ \ \ \ \ \ F_{1}$ & $\ \ \ \ \ \ \ \
F_{2} $ & $\ \ \ \ \ \ \ \ \ \ F_{3}$ \\ \hline
Fixed Points & $\left( 0,0,z_{3},z_{4},z_{5}\right) $ & $\left(
0,z_{2},0,z_{4},z_{5}\right) $ & $\left( 0,z_{2},z_{3},0,z_{5}\right) $ \\
\hline
${\bf B}_{a}={\bf CP}^{2}$ & ${\bf B}_{12}=\left\{ z_{3},z_{4},z_{5}\right\}
$ & ${\bf B}_{23}=\left\{ z_{2},z_{4},z_{5}\right\} $ & ${\bf B}%
_{31}=\left\{ z_{2},z_{3},z_{5}\right\} $ \\ \hline
${\cal S}_{a}$ Fibers &
\begin{tabular}{l}
$u_{1}+w_{12}+\frac{w_{12}^{5}}{u_{1}}=0$ \\
$u_{1}=z_{1}^{5}$, $u_{2}=z_{2}^{5}$, \\
$w_{13}=z_{1}z_{3}$.
\end{tabular}
&
\begin{tabular}{l}
$u_{1}+w_{13}+\frac{w_{13}^{5}}{u_{1}}=0$ \\
$u_{1}=z_{1}^{5}$, $u_{3}=z_{3}^{5}$, \\
$w_{12}=z_{1}z_{2}$.
\end{tabular}
&
\begin{tabular}{l}
$u_{1}+w_{14}+\frac{w_{14}^{5}}{u_{1}}=0$ \\
$u_{1}=z_{1}^{5}$, $u_{4}=z_{4}^{5}$, \\
$w_{14}=z_{1}z_{4}$.
\end{tabular}
\\ \hline
\end{tabular}
\end{equation}

\bigskip

\qquad Having the above features in mind, the singular representations of
the NC quintic may be obtained by starting from the regular representation
eqs(5.16) and taking the appropriate limits. For the ${\cal S}_{1}$ singular
space for instance, the fields moduli are obtained from eqs(5.16) by setting
$q_{i}^{1}=q_{i}^{2}=0$ \ and taking the zero limit of $\ z_{1}$ and $z_{2}$. We have for the fiber ${\cal S}_{1}$,

\begin{eqnarray}
\Phi _{1} &=&\ \left( {{Q}\otimes }I_{id}\otimes I_{id}\right) \
\lim_{z_{1}\rightarrow 0}\ {z_{1}},  \nonumber \\
\Phi _{2} &=&\ \left( {{Q^{-1}}}\otimes I_{id}\otimes I_{id}\right) \
\lim_{z_{1}\rightarrow 0}\ {z_{2}}, \\
F_{1} &=&P\otimes Q^{\eta _{12}}\otimes Q^{\eta _{13}},  \nonumber
\end{eqnarray}

while for ${\bf B}_{1}$, we have:

\begin{eqnarray}
\Phi _{3} &=&z_{3}\ \left( I_{id}{\otimes }I_{id}\otimes I_{id}\right) ,
\nonumber \\
\Phi _{4} &=&z_{4}\ \left( I_{id}{\otimes }I_{id}\otimes I_{id}\right) ,
\nonumber \\
\Phi _{5} &=&z_{5}\ \left( I_{id}{\otimes }I_{id}\otimes I_{id}\right) , \\
F_{2} &=&Q^{\eta _{21}}\otimes P\otimes Q^{\eta _{23}},  \nonumber \\
F_{3} &=&Q^{\eta _{31}}\otimes Q^{\eta _{32}}\otimes P.  \nonumber
\end{eqnarray}

As one sees, once the limit to the singular point is taken, the non
vanishing matrix field moduli $\Phi _{i}$,\ $i=3,4,5$ \ are proportional to
the identity $\ \left( I_{id}{\otimes }I_{id}\otimes I_{id}\right) ={\bf I}_{%
{\cal D}\left( G\right) }$\ of the group representation \ ${\cal D}\left(
G\right) $; i.e \ $\Phi _{i}=z_{i}\otimes {\bf I}_{{\cal D}\left( G\right) }$. This a very remarkable feature at singularity which have an algebraic and
a brane interpretations.

{\it Fractional branes on \ }${\cal R^{\prime }}/{\bf Z}_{5}$

To fix the ideas, let start from the $D9$ brane of type $IIB$ string wrapped
on the quintic $Q$; and whose coordinates are denoted as $\left( x^{\mu };%
\text{ }y_{1}\text{, }y_{2}\text{, }y_{3}\right) $ with $x^{\mu }=\left(
x^{0},\text{ }x^{1}\text{, }x^{2},\text{ }x^{3}\right) $ the longitudinal
non compact variables of ( representing mainly a $D3$ brane embedded in $D9$) and the $y_{i}$'s
are the compact transverse coordinates of the wrapped $D9$ branes. In other
words, the $D9$ brane may naively be thought of as; $D9\sim D3\times {\cal Q}
$. In the coordinate patch $z_{5}=z_{4}=1$ , the $y$ coordinates may be
imagined as related to those of the quintic as,
\begin{eqnarray}
y_{1} &=&z_{1},  \nonumber \\
y_{2} &=&z_{2}, \\
y_{3} &=&\frac{-a_{0}z_{1}z_{2}z_{3}}{\left(
2+z_{1}^{5}+z_{2}^{5}+z_{3}^{5}\right) },  \nonumber
\end{eqnarray}

In the case where the orbifold of the quintic is given  by $Q^{\left[ 1\right] }$ and
torsion taken into account, the wrapped $D9$ becomes a non commutative brane
generated by the algebra eqs(6.1) or equivalently by eqs(5.1). At the fixed
point of the $Q^{\left[ 1\right] }$ orbifold, where a real two cycle shrinks
to zero, the NC wrapped $D9$ \ becomes a priori a NC wrapped $D7$ described
by;

\begin{eqnarray}
\Phi _{3} &=&z_{3}\ {I}_{{\cal D}\left( {\bf Z}_{5}\right) },\qquad \Phi
_{4}=z_{4}\ {I}_{{\cal D}\left( {\bf Z}_{5}\right) },\qquad \Phi _{5}=z_{5}\
{I}_{{\cal D}\left( {\bf Z}_{5}\right) },  \nonumber \\
F_{1} &=&P\text{,}\qquad F_{2}=Q^{\eta _{21}},\qquad F_{3}=Q^{\eta _{31}}%
\text{.}
\end{eqnarray}

while the singular modes at the orbifold point are carried by the ${Q}$
operator and its inverse ${{Q^{-1}}}$ as shown on the following eqs.

\begin{equation}
\Phi _{1}=\ {Q}\text{ }\ \lim_{z_{1}\rightarrow 0}\ {z_{1}},\qquad \Phi
_{2}=\ {{Q^{-1}}}\ \lim_{z_{2}\rightarrow 0}\ {z_{2}}.
\end{equation}

However due to the complete reducibility property of ${I}_{{\cal D}\left(
{\bf Z}_{5}\right) }$ \ namely \ ${I}_{{\cal D}\left( {\bf Z}_{5}\right)
}=\sum_{n=1}^{5}\Pi _{n}$,\ eqs(7.19) shows that they describe rather a set
of five commuting Euclidean wrapped $D7$\ branes parameterized as
\begin{equation}
\Phi _{3,n}=z_{3}\ \Pi _{n},\qquad \Phi _{4,n}=z_{4}\ \Pi _{n},\qquad \Phi
_{5}=z_{5,n}\ \Pi _{n}\text{.}
\end{equation}

Therefore at the orbifold point of ${\cal R^{\prime }}/{\bf Z}_{5}$, we have
five fractional wrapped $D7$ branes. Moreover since near the singularity the
$\Phi _{1}$ and $\Phi _{2}$ singular operator may also be split as
\[
\Phi _{1}=\left( \lim_{z_{1}\rightarrow 0}\ {z_{1}}\right)
\sum_{n=1}^{5}a_{n}^{+},\qquad \Phi _{2}=\left( \lim_{z_{2}\rightarrow 0}\ {%
z_{2}}\right) \sum_{n=1}^{5}a_{n}^{-},
\]

there are also massless modes $\phi _{1,n}\sim a_{n}^{+}$\ and $\phi
_{2,n}\sim a_{n}^{-}$\ \ living on the wrapped $D7$ branes; they are
propagating modes traveling  between $\Phi _{i,n}$\ and $\Phi _{i,n\pm 1}$\ fractional
branes. The quiver diagram representing the fractional branes is also a $%
\Delta _{5}$\ pentagon with wrapped $D7$ siting at the vertices and massless
modes represented by the links.

\qquad A similar analysis to that we developed above may be also written down for the
${\bf B}_{2}\bowtie $\ ${\cal S}_{2}$\ and \ ${\bf B}_{3}\bowtie $\ ${\cal S}%
_{3}$ quintic fibrations. The five links joining the neighboring \ $\pi _{n}$%
\ nodes describe massless fields of the effective field theory on the $D4$
branes at singularity. For each fibration, there are $(2\times 5)$ massless
complex fields which we denote as $\chi _{a,k_{a}}\sim a_{k_{a}}^{+}$ and $%
\psi _{a,k_{a}}\sim a_{k_{a}}^{-}$. In particular at $z_{1}=z_{2}=z_{3}=0$,
we have therefore triplets of massless fields as shown on table eq(7.24) where we give the fields spectrum on the fractional wrapped $D7$
branes on the ${\bf B}_{a}$\ basis at the various singularities.

{\it Fields Spectrum}\newline

\begin{equation}
\begin{tabular}{|c|c|c|}
\hline
Chiral fields $\rightarrow $ & Massless complex fields & Number of massless
fields \\[2mm] \hline
Singularity ${\cal S}_{1}$ & $\chi _{1,k_{1}}$ , $\psi _{1,k_{1}}$ & $%
2\times 5$ \\[2mm] \hline
Singularity ${\cal S}_{2}$ & $\chi _{2,k_{2}}$ , $\psi \ _{2,k_{2}}$ & $%
2\times 5$ \\[2mm] \hline
Singularity ${\cal S}_{3}$ & $\chi \ _{3,k_{3}}$ , $\psi \ _{3,k_{3}}$ & $%
2\times 5$ \\[2mm] \hline
Locus $z_{1}=0$ & $\chi _{1,k_{1}},\chi \ _{2,k_{2}},\chi \ _{3,k_{3}}$ & $%
3\times 5$ \\[2mm] \hline
\end{tabular}
\end{equation}

Such analysis can also be extended to the case where the orbifold of the
quintic is realized as ${\cal Q}^{\left[ 2\right] }\sim {\cal R^{\prime
\prime }}/{\bf Z}_{5}^{2}$.

\subsection{ Fractional Branes on ${\cal Q}^{\left[ 2\right] }$}

\qquad There are three manifest fibrations of the commutative quintic
orbifold ${\cal Q}^{\left[ 2\right] }$ depending on the nature of the ${\bf Z%
}_{5}^{2}$ subgroups of ${\bf Z}_{5}^{3}$   one is considering. Let us
describe them briefly here below

\subsubsection{ Fibration ${\cal Q}^{\left[ 2\right] }\simeq CP^{1}\bowtie
{\cal S}_{12}$}

\qquad Here the ${\bf Z}_{5}^{2}$ orbifold subsymmetry of the ${\bf Z}%
_{5}^{2}$ group has the following charges
\begin{eqnarray}
q_{i}^{1} &=&\left( 1,-1,0,0,0\right) ,  \nonumber \\
q_{i}^{2} &=&\left( 1,0,-1,0,0\right) .
\end{eqnarray}

The $q_{i}^{3}$ charges of the third factor may be thought as being set
equal to zero. \ As such the ${\bf B}_{12}$ basis of the fibration is
parameterized by the $z_{4}$ and $z_{5}$\ coordinates; while the codimension
one fiber ${\cal S}_{12}$ is given by the following complex surface with an $%
{\bf A}_{4}$ singularity,
\begin{equation}
u+v+t+\frac{t^{5}}{uv}=0,
\end{equation}

Like for the ${\bf B}_{a}\bowtie $\ ${\cal S}_{a}$\ fibrations, the\ ${\bf B}%
_{ab}$\ base manifolds are parameterized by those holomorphic coordinates $%
z_{i}$ with $q_{i}^{a}=0$ while the ${\cal S}_{ab}$ fibers are parameterized
by the complex variables with non zero $q_{i}^{a}$ charges with fixed points
at the origin.

\subsubsection{Fibration ${\cal Q}^{\left[ 2\right] }\simeq CP^{1}\bowtie
{\cal S}_{23}$}

\qquad In this fibration, the ${\bf Z}_{5}^{2}$ orbifold subgroup has the
following charges
\begin{eqnarray}
q_{i}^{2} &=&\left( 1,0,-1,0,0\right) ,  \nonumber \\
q_{i}^{3} &=&\left( 1,0,0,-1,0\right) .
\end{eqnarray}

The ${\bf B}_{23}$ basis and the the ${\cal S}_{23}$\ fiber of the quintic
are respectively parameterized by $\left( z_{2},z_{5}\right) $\ and $\left(
z_{1},\ z_{3},\ z_{4}\right) $ complex variables. Locally ${\bf B}_{23}\sim
{\bf C}$ while the ${\cal S}_{23}$ singular surface is of type eq(7.24) with
an ${\bf A}_{4}$ singularity at the origin.

\subsubsection{Fibration ${\cal Q}^{\left[ 2\right] }\simeq CP^{1}\bowtie
{\cal S}_{31}$}

\qquad In this case, the Calabi-Yau charges of the ${\bf Z}_{5}^{2}$
orbifold subgroup are;
\begin{eqnarray}
q_{i}^{1} &=&\left( 1,-1,0,0,0\right) ,  \nonumber \\
q_{i}^{3} &=&\left( 1,0,0,-1,0\right) .
\end{eqnarray}

The base is parameterized by the invariant $\left( \ z_{3},z_{5}\right) $\
coordinates, while the ${\cal S}_{31}$ fiber has a singularity at $%
z_{1}=z_{2}=z_{4}=0$. \ Its complex equation \ $u_{1}+u_{2}+t+\frac{t^{5}}{%
u_{1}u_{2}}=0$\ \ has an ${\bf A}_{4}$ singularity at $\ u_{1}=u_{2}=t=0$.

\qquad The main features of the various ${{\bf B}_{\left( ab\right) }}%
\bowtie {\cal S}_{\left( ab\right) }$ fibrations are collected in this table

\bigskip

\begin{equation}
\begin{tabular}{|l|l|l|l|}
\hline
$F_{a}\otimes F_{b}$ & $F_{1}\otimes F_{2}$ & $F_{2}\otimes F_{3}$ & $%
F_{3}\otimes F_{1}$ \\ \hline
Fix Points & $\left( 0,0,0,z_{4},z_{5}\right) $ & $\left(
0,z_{2},0,0,z_{5}\right) $ & $\left( 0,0,z_{3},0,z_{5}\right) $ \\ \hline
${\bf B}_{ab}$ & ${\bf CP}^{1}=\left\{ z_{4},z_{5}\right\} $ & ${\bf CP}%
^{1}=\left\{ z_{2},z_{5}\right\} $ & ${\bf CP}^{1}=\left\{
z_{3},z_{5}\right\} $ \\ \hline
\begin{tabular}{l}
Fibers \\
${\cal S}_{ab}$%
\end{tabular}
&
\begin{tabular}{l}
$u_{1}+u_{2}+t+\frac{t^{5}}{u_{1}u_{2}}=0$ \\
$u_{1}=z_{1}^{5}$, $u_{2}=z_{2}^{5}$, \\
$t=z_{1}z_{2}z_{3}$.
\end{tabular}
&
\begin{tabular}{l}
$u_{1}+u_{3}+v+\frac{v^{5}}{u_{1}u_{2}}=0$ \\
$u_{1}=z_{1}^{5}$, $u_{3}=z_{3}^{5}$, \\
$v=z_{1}z_{3}z_{4}$%
\end{tabular}
&
\begin{tabular}{l}
$u_{1}+u_{4}+w+\frac{w^{5}}{u_{1}u_{4}}=0$ \\
$u_{1}=z_{1}^{5}$, $u_{4}=z_{4}^{5}$, \\
$w=z_{1}z_{2}z_{4}$.
\end{tabular}
\\ \hline
Fields & $\chi _{1,\left( k_{1},k_{2}\right) }$ $,$ $\psi _{1,\left(
k_{1},k_{2}\right) }$ & $\chi _{2,\left( k_{2},k_{3}\right) }$ $,$ $\psi
_{2,\left( k_{2},k_{3}\right) }$ & $\chi _{3,\left( k_{1},k_{3}\right) }$ $,$
$\psi _{3,\left( k_{1},k_{3}\right) }$ \\ \hline
\end{tabular}
\end{equation}

\bigskip

\qquad The singular representations of non commutative ${\cal Q}^{nc\left[ 2%
\right] }$ \ may be obtained by starting from the regular representation
eqs(5.16) and taking the appropriate limits. For the ${\cal S}_{12}$
singular space, the fields moduli are obtained from eqs(5.16) by setting $%
q_{i}^{1}=q_{i}^{2}=0$ \ and taking $\ z_{1}$, $z_{2}$\ and $z_{3}$ to zero.
For ${\cal S}_{12}$ fiber,we get:

\begin{eqnarray}
\Phi _{1} &=&\ \left( {{Q}\otimes }I_{id}\otimes I_{id}\right) \
\lim_{z_{1}\rightarrow 0}\ {z_{1}},  \nonumber \\
\Phi _{2} &=&\ \left( {{Q^{-1}}}\otimes I_{id}\otimes I_{id}\right) \
\lim_{z_{2}\rightarrow 0}\ {z_{2}},  \nonumber \\
\Phi _{3} &=&\ (I_{id}\text{ }{\otimes {Q^{-1}}\otimes }I_{id})\
\lim_{z_{3}\rightarrow 0}\ {z_{3}}, \\
F_{1} &=&P\otimes Q^{\eta _{12}}\otimes Q^{\eta _{13}},  \nonumber \\
F_{2} &=&Q^{\eta _{21}}\otimes P\otimes Q^{\eta _{23}},  \nonumber
\end{eqnarray}

while for ${\bf B}_{12}$, we have:

\begin{eqnarray}
\Phi _{4} &=&z_{4}\ \otimes {\bf I}_{{\cal D}\left( G\right) },  \nonumber \\
\Phi _{5} &=&z_{5}\ \otimes {\bf I}_{{\cal D}\left( G\right) }, \\
F_{3} &=&Q^{\eta _{31}}\otimes Q^{\eta _{32}}\otimes P.  \nonumber
\end{eqnarray}

Once the limit to the singular point is taken, the non vanishing matrix
field moduli $\Phi _{i}$,\ $i=4,5$ \ are proportional to the identity ${\bf I%
}_{{\cal D}\left( G\right) }$\ of the group representation \ ${\cal D}\left(
G\right) $; but as before, this property reflects just the existence of
fractional $D$ branes at the orbifold point $z_{1}=z_{2}=z_{3}=0$.

{\it Fractional branes on \ }${\cal Q}^{\left[ 2\right] }$

\qquad Let us reconsider the example of the $D9$ brane of type $IIB$ string
wrapped on ${\cal Q}^{\left[ 2\right] }$; with local  coordinates $%
\left( x^{\mu };\text{ }y_{1}\text{, }...\text{, }y_{3}\right) $;. the $%
x^{\mu }$'s being the longitudinal non compact coordinates of the $D3$ part
of $D9,$ while the $y$ compact coordinates are as in eqs(7.20).

In case where torsions of ${\cal Q}^{\left[ 2\right] }$ are taken into
account, the wrapped $D9$ on \ ${\cal Q}^{\left[ 2\right] }$\ becomes a non
commutative brane generated by the algebra eqs(5.1) where the calabi-Yau
charges are chosen as in eqs(7.23). At the fixed point of the ${\cal Q}^{%
\left[ 2\right] }$ orbifold, where a real four cycle shrinks to zero, the NC
wrapped $D9$ give rise to twenty five fractional wrapped $D5$ branes on $%
{\bf B}_{12}\sim {\bf CP}^{1}$. The transverse coordinates of these
fractional branes are given by;

\begin{eqnarray}
\Phi _{4;n,m} &=&x_{4}\ \Pi _{n,m},  \nonumber \\
\Phi _{5;n,m} &=&x_{5}\ \Pi _{n,m},
\end{eqnarray}

where $\Pi _{n,m}$\ are the projectors on the ${\cal D}\left( {\bf Z}%
_{5}^{2}\right) $ representation states. The singular modes at the orbifold
point are carried by the ${Q\otimes Q}$, ${{Q^{-1}}}\otimes I_{id}$ and $%
I_{id}\otimes {{Q^{-1}}}$ operators of ${\cal D}\left( {\bf Z}%
_{5}^{2}\right) $. Moreover since near the singularity the $\Phi _{1}$, $%
\Phi _{2}$ and $\Phi _{3}$ operators may also be split as
\begin{eqnarray}
\Phi _{1} &=&\left( \lim_{z_{1}\rightarrow 0}\ {x_{1}}\right)
\sum_{n_{1},n_{2}=1}^{5}A_{n_{1}}^{+}A_{n_{2}}^{+},  \nonumber \\
\Phi _{2} &=&\left( \lim_{z_{2}\rightarrow 0}\ {x_{2}}\right)
\sum_{n_{1},n_{2}=1}^{5}A_{n_{1}}^{-}\Pi _{n_{2}}, \\
\Phi _{3} &=&\left( \lim_{z_{3}\rightarrow 0}\ {x_{3}}\right)
\sum_{n_{1},n_{2}=1}^{5}\Pi _{n_{1}}A_{n_{2}}^{-},  \nonumber
\end{eqnarray}

where $A_{n_{a}}^{\pm }$\ were introduced in subsection eqs(4.30), one gets
as a consequence $3\times 25$ massless modes $\phi _{1;n_{1},n_{2}}\sim
A_{n_{1}}^{+}A_{n_{2}}^{+}$, $\phi _{2;n_{1},n_{2}}\sim A_{n_{1}}^{-}\Pi
_{n_{2}}$\ \ and $\phi _{2;n_{1},n_{2}}\sim \Pi _{n_{1}}A_{n_{2}}^{-}$\ \
living on the $D5$ branes wrapped on ${\bf CP}^{1}$; they are propagators
between the $\Phi _{i;n_{1},n_{2}}$\ and $\Phi _{i,n_{1}\pm 1,n_{2}\pm 1}$\
fractional $D5$ branes. The quiver diagram representing the fractional $D5$
branes wrapped on ${\bf CP}^{1}$ is also a $\Delta _{5}\times \Delta _{5}$\
polygon with $D5$ branes sitting at the vertices and the $\phi
_{a;n_{1},n_{2}}$ massless modes propagating along the links, see figure 4.

\begin{figure}[tbh]
\begin{center}
\epsfig{file=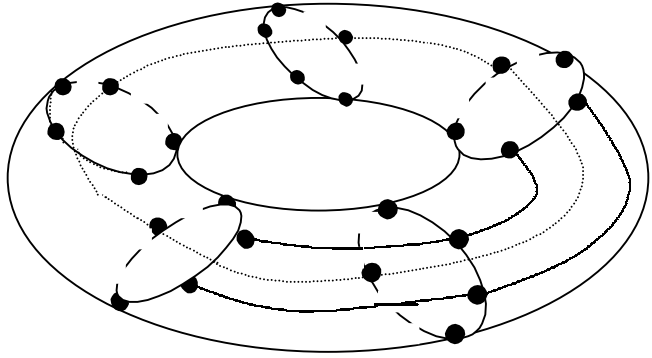}
\end{center}
\caption{{\protect\small {\it Here is represented the 25 vertices of the
quiver diagram of the fractional $D$ branes at $x_{1}=x_{2}=x_{3}=0$
singularity of the ${\bf Z}^{2}$ orbifold subsymmetry of the quintic. The
links between the different vertices represents the massless modes on the
wrapped $D5$ brane on the base ${\bf B}_{12} = {\bf CP}^{1}$ }}}
\end{figure}

\qquad An analogous analysis may be also made for the ${\bf B}_{23}\bowtie $%
\ ${\cal S}_{23}$\ and \ ${\bf B}_{31}\bowtie $\ ${\cal S}_{31}$ fibrations
of the quintic; we get similar results.

\section{Discussion and Conclusion}

\qquad In this paper we have studied \ non commutative geometry versus
discrete torsion and given an explicit computation of the spectrum of
fractional $D$ branes at singularities of Calabi-Yau hypersurface orbifolds $%
{\cal O}={\cal M}/G$. First we have drawn the general setting of this kind
of NC geometry and given links with string compactifications on orbifolds
with discrete torsion.

Taking the quintic as an illustrating example, we have worked out
general and explicit solutions for ${\cal Q}^{nc}$, the non commutative
extension of the quintic using first a constrained method and then a more
involved one baptized as the CP algebra and denoted as ${\cal Q}\boxtimes
{\cal A}_{{\cal D}\left( G\right) }$. The latter is a fibration whose base
is just the commutative orbifold ${\cal O}$, which is also the centre of $%
{\cal O}\boxtimes {\cal A}_{{\cal D}\left( G\right) }$, and its fiber is
given by ${\cal A}_{{\cal D}\left( G\right) }$, the algebra of the orbifold
group representation ${\cal D}\left( G\right)$. For the case of the quintic,
${\cal Q}={\cal O}$, where the coset symmetry $G_{\left[ \alpha \right] }$   is a subgroup  of
${\bf Z}_{5}^{3}$; that is $G_{\left[ 1\right] }={\bf Z}_{5}$\ or $G_{\left[
2\right] }={\bf Z}_{5}^{2}$,\ and in presence of a torsion matrix $\eta
_{ab} $, we have shown that NC geometry is governed by the following features;

(i) The matrix $\eta _{ab}$\ should be a non symmetric  invertible matrix belonging to $SL\left( 3,{\bf Z}\right) $.

(ii) In addition to the classical symmetries of ${\cal Q}$ namely invariance
under $S_{5}$ permutations of the complex variables and the ${\bf Z}_{5}^{3}$
invariance acting as $z_{i}\longrightarrow z_{i}$ $\omega ^{q_{i}^{a}}$,
there is an extra hidden symmetry ${\bf Z}_{5}^{3}$ induced by torsion, and
acting as $z_{i}\longrightarrow z_{i}$ $\omega ^{p_{i}^{a}}$ with $p_{i}^{a}$%
\ charges given by
\begin{equation}
p_{i}^{a}=\eta _{ab}\text{ }q_{i}^{b}\text{.}
\end{equation}

Note that for $\eta _{ab}=0$\ ( no torsion), there is no $p_{i}^{a}$'s.

(iii) To each abelian factor ${\bf Z}_{5}=\left\{ \omega _{i};\omega
_{i}^{5}=I_{id}\right\} $\ of the geometric symmetry ${\bf Z}_{5}^{3}$\ of
the quintic, it is associated an automorphism symmetry ${\bf Z}_{5}^{aut}$
rotating the $\omega _{i}$ elements of ${\bf Z}_{5}$. These two groups do
not commute; ${\bf Z}_{5}\ast {\bf Z}_{5}^{aut}\neq {\bf Z}_{5}^{aut}\ast
{\bf Z}_{5}^{aut}$ and were shown to describe a Fuzzy torus ${\cal T}%
_{\kappa }^{2}$\ with $\kappa $, a given fifth root of unity, say $\kappa
=\omega =\exp i\frac{2\pi }{5}$. This feature shows that the effective coset symmetry of the
orbifold of the quintic is $\left( {\bf Z}_{5}\ast {\bf Z}_{5}^{aut}\right)
^{3}$\ rather than ${\bf Z}_{5}^{3}$. Moreover the solutions for the ${\bf Z}%
_{i}$ generators of the NC quintic are shown to depend on the ${\cal D}%
\left( \left( {\bf Z}_{5}\ast {\bf Z}_{5}^{aut}\right) ^{3}\right) $ group
realization and can be thought of as describing a five dimensional Fuzzy
torus ${\cal T}_{\beta _{ij}}^{5}$, whose $\beta _{ij}$ cocycle parameters
are given by
\begin{equation}
\beta _{ij}=\omega ^{\sum_{a=1}^{3}\left(
p_{i}^{a}q_{j}^{a}-p_{j}^{a}q_{i}^{a}\right) }
\end{equation}

Such results extend to any complex $n$ dimension hypersurface ${\cal H}_{n}$%
, $z_{1}^{n+2}+...+z_{n+2}^{n+2}+a_{0}\prod_{i=1}^{n+2}z_{i}=0$, \ with
orbifold group ${\bf Z}_{n+2}^{n}$. In this case the full symmetry is $%
\left( {\bf Z}_{n+2}\ast {\bf Z}_{n+2}^{aut}\right) ^{\otimes n}$, the
torsion matrix $\eta _{ab}$\ should be invertible and non symmetric matrix
belonging to $SL\left( n,{\bf Z}\right) $. The corresponding non commutative
geometry is mainly given by a real $\left( n+2\right) $ dimensional Fuzzy
torus ${\cal T}_{\beta _{ij}}^{n+2}$ with $\beta _{ij}$ cocycles as
\begin{equation}
\beta _{ij}=\prod_{a=1}^{3}\exp \left[ -i\frac{2\pi }{n}\left( \eta
_{ab}-\eta _{ba}\right) q_{i}^{a}q_{j}^{b}\right] \text{,}
\end{equation}

where now the $n$ vectors ${\bf q}^{a}=\left( q_{i}^{a}\right) $\ have $%
\left( n+2\right) $ integer entries; they satisfy the Calabi-Yau condition $%
\sum_{i=1}^{n+2}q_{i}^{a}=0$ \ and, for simple and explicit applications,
may be chosen as,
\begin{eqnarray}
{\bf Z}_{n+2} &:&\qquad {\bf q}^{1}=\left( 1,-1,0,0,...,0,0,0\right) ,\qquad
\qquad \qquad \qquad \left( 1\right)  \nonumber \\
{\bf Z}_{n+2} &:&\qquad {\bf q}^{2}=\left( 1,0,-1,0,...,0,0,0\right) ,\qquad
\qquad \qquad \qquad \left( 2\right)  \nonumber \\
&&\qquad ... \\
{\bf Z}_{n+2} &:&\qquad {\bf q}^{n-1}=\left( 1,0,0,0,...,-1,0,0\right)
,\qquad \qquad \qquad \left( n-1\right)  \nonumber \\
{\bf Z}_{n+2} &:&\qquad {\bf q}^{n}=\left( 1,0,0,0,...,0,-1,0\right) .\qquad
\qquad \qquad \qquad \left( n\right)  \nonumber
\end{eqnarray}

\bigskip

Because of the richer symmetry of ${\cal H}_{n}=\left\{
z_{1}^{n+2}+...+z_{n+2}^{n+2}+a_{0}\prod_{i=1}^{n+2}z_{i}=0\right\} $, the
study of fractional branes at orbifold points of ${\cal H}_{n}/G_{\left[
\alpha \right] }$ is also rich and depends on the orbifold subgroups $G_{%
\left[ \alpha \right] }$ of ${\bf Z}_{n+2}^{n}$  one is considering.\ There are several
subgroups in ${\bf Z}_{n+2}^{n}$, the natural ones are the manifest factors
$G_{\left[ \alpha \right] }={\bf Z}_{n+2}^{\alpha }$, with $1\leq \alpha
\leq \left( n-1\right) $. Let us give some comments regarding this point,
but forget for a while about string applications by letting $n$ to be a
generic positive integer greater than two and suppose that we have a $p$%
-brane, $p>n$, wrapping the ${\cal H}_{n}$\ compact manifold.

{\it Fractional branes on} ${\cal H}_{n}/{\bf Z}_{n+2}$

\qquad Since there are $n$ manifest ${\bf Z}_{n+2}$\ subsymmetries of type
eqs(8.4-1,8.4-n) in ${\bf Z}_{n+2}^{n}$, we have $n$ classes of fractional $%
\left( p-2\right) $-branes at the $n$ kinds of singularities $%
z_{1}=z_{a+1}=0 $. Extending the analysis we have made for the quintic to
generic ${\cal H}_{n}$'s by thinking about ${\cal H}_{n}/{\bf Z}_{n+2}$\ as
a fiber bundle ${\bf B}_{a}\bowtie {\cal S}_{a}$ of base ${\bf B}_{a}={\bf CP%
}^{n-1}$ and a fiber ${\cal S}_{a}$ given by the complex curve $u+v+\frac{%
v^{n+2}}{u}=0$ with an ${\bf A}_{n-1}$ singularity at the origin $u=v=0$, we
will have $d_{1}$ fractional branes at each orbifold point, with $d_{1}=\dim
\left( {\cal D}\left( {\bf Z}_{n+2}\right) \right) $, and $2n$ massless
modes living on the $\left( p-2\right) $ branes. Points in this NC geometry
are represented by polygons $\Delta _{n+2}$ with $\left( n+2\right) $\
vertices and $\left( n+2\right) $ edges.

{\it Fractional branes on} ${\cal H}_{n}/{\bf Z}_{n+2}^{2}$

\qquad Before giving the result concerning ${\cal H}_{n}/{\bf Z}_{n+2}^{2}$,
note that the action of ${\bf Z}_{n+2}^{2}$ on the $z_{i}$ variables is
additive; that is if one performs two successive ${\bf Z}_{n+2}$ actions
with different Calabi-Yau charges, say $q_{i}^{a}$ and $q_{i}^{b}$
respectively, then the total ${\bf Z}_{n+2}^{2}$\ action has the charge $%
q_{i}^{a}+q_{i}^{b}$,
\begin{eqnarray}
{\bf Z}_{n+2} &:&\qquad z_{i}\longrightarrow z_{i}\omega ^{q_{i}^{a}},
\nonumber \\
{\bf Z}_{n+2} &:&\qquad z_{i}\longrightarrow z_{i}\omega ^{q_{i}^{b}}, \\
{\bf Z}_{n+2}^{2} &:&\qquad z_{i}\longrightarrow z_{i}\omega ^{\left(
q_{i}^{a}+q_{i}^{b}\right) }.  \nonumber
\end{eqnarray}

From these relations, one sees that all happen as if one has only the ${\bf Z%
}_{n+2}^{\prime }$ diagonal subsymmetry of ${\bf Z}_{n+2}^{2}$ with $\left(
q_{i}^{a}+q_{i}^{b}\right) $ charge; the off diagonal subsymmetry, with $%
\left( q_{i}^{a}-q_{i}^{b}\right) $, is not taken into account. For the
example of the ${\bf Z}_{n+2}^{2}$ symmetry associated to eq (8.4-1) and
eq(8.4-2), the diagonal ${\bf Z}_{n+2}^{\prime }$ symmetry has the following
vector charge
\begin{equation}
{\bf Z}_{n+2}^{2}:\qquad {\bf q}^{1}+{\bf q}^{2}=\left(
2,-1,-1,0,...,0,0,0\right) ,
\end{equation}

and fix points at $z_{1}=z_{2}=z_{3}=0$ \ and more generally at $%
z_{1}=z_{a+1}=z_{b+1}=0$. This property shows that:

(a) the fibration ${\bf B}_{ab}\bowtie {\cal S}_{ab}$ of the manifold ${\cal %
H}_{n}$ has as a base ${\bf B}_{ab}={\bf CP}^{n-2}$ and as a fiber ${\cal S}%
_{ab}$ the complex surface $u+v+t+\frac{t^{n+2}}{uv}=0$ with an ${\bf A}%
_{n-1}$ singularity at the origin $u=v=t=0$.

(b) there are $\frac{n\left( n-1\right) }{2}$\ different fibrations of $%
{\cal H}_{n}$ and so there are $\frac{n\left( n-1\right) }{2}$\ classes of
fractional $\left( p-4\right) $-branes at the $z_{1}=z_{a+1}=z_{b+1}=0$ \
singularities. Each class contains $d_{1}d_{2}$ fractional $\left(
p-4\right) $-branes. Using a similar analysis to that of subsections 7.2-3,
one can check that there are $2\left( d_{1}+d_{2}\right) $\ massless modes
living on theses branes, with $\ d_{1}+d_{2}=\dim {\cal D}\left( {\bf Z}%
_{n+2}^{2}\right) $. Points in this NC geometry are given by the crossed
product of two polygons $\Delta _{d_{1}}\times \Delta _{d_{2}}$.

{\it Fractional branes on} ${\cal H}_{n}/{\bf Z}_{n+2}^{k}$

\qquad Extending the above reasoning to ${\cal H}_{n}/{\bf Z}_{n+2}^{k}$,
the orbifold symmetry ${\bf Z}_{n+2}^{k}$\ with charges ${\bf q}^{a_{1}}$, $%
...$, ${\bf q}^{a_{k}}$\ acts in practice through it diagonal ${\bf Z}_{n+2}$
subgroup with a Calabi-Yau charge $Q_{\left( a_{1},...,a_{k}\right) }={\bf q}%
^{a_{1}}+...+{\bf q}^{a_{k}}$; it has fix points located at $%
z_{1}=z_{a_{1}+1}=...=z_{a_{k}+1}=0$. For the example of the leading $k$
symmetries of eqs(8.4), the corresponding diagonal symmetry charge $%
Q_{\left( 1,...,k\right) }$ is given by;
\begin{equation}
{\bf Z}_{n+2}^{k}:\qquad Q_{\left( 1,...,k\right) }=\left(
k,-1,-1,...,-1,0,...,0,0\right) ;
\end{equation}

where the first $\left( k+1\right) $ entries of $Q_{\left( 1,...,k\right) }$
are non zero while the remaining ones are all of them equal to zero. A
simple counting shows that there are $\frac{n!}{\left( n-k\right) !k!}$
possible subgroups of this kind in ${\bf Z}_{n+2}^{n}.$

\qquad The fibration ${\bf B}_{\left( a_{1}...a_{k}\right) }\bowtie {\cal S}%
_{\left( a_{1}...a_{k}\right) }$ of the manifold ${\cal H}_{n}$ has as a
base ${\bf B}_{\left( a_{1}...a_{k}\right) }={\bf CP}^{n-k}$ and as a fiber $%
{\cal S}_{\left( a_{1}...a_{k}\right) }$ the complex dimension $k$
hypersurface $u_{1}+...+u_{k}+t+\frac{t^{n+2}}{\varrho }=0$, $\varrho
=\prod_{j=1}^{k}u_{j}$, with an ${\bf A}_{n-1}$ singularity at $%
u_{1}=...=u_{k}=t=0$. An other result is that there are $\frac{n!}{\left(
n-k\right) !k!}$\ different fibrations of ${\cal H}_{n}$ and so there are $%
\frac{n!}{\left( n-k\right) !k!}$\ classes of fractional $\left( p-2k\right)
$-branes at the $u_{1}=...=u_{k}=t=0$ singularities. Each class contains $%
\left( \prod_{j=1}^{k}d_{j}\right) $ fractional $\left( p-2k\right) $-branes
$k\sum_{i=1}^{k}d_{i}$\ \ massless modes living on theses branes, with $\
\sum_{i=1}^{k}d_{i}$ being the dimension of the ${\cal D}\left( {\bf Z}%
_{n+2}^{k}\right) $ representation group of ${\bf Z}_{n+2}^{k}$.

\qquad The analysis we made for the quintic extends to all homogeneous
hypersurfaces and the results we have obtained are also valid. For the
special case $n=4$, we recover known results on orbifolds of the ${\bf K}%
_{3} $ surface. In future work, we will discuss the case NC local Calabi-Yau
manifolds embedded in NC toric varieties.

{\bf Acknowledgments}\newline
This research work is supported by DFG/CNCPRST 445 Mar 113/5/0 contract between University of Munich, Germany, and University of Rabat, Morocco. I am very grateful to Julius Wess for kind hospitality at Munich University, to  DFG program which  supported  my  stay in Germany and to CNR Morocco for covering my Air faire. I thank I. Benkaddour, C. Blooman for discussions and  A.Belhaj and  M.Bennai for earlier collaboration on this subject. Many thanks to Mrs
M.Kurzinger and to the team of the theoretical Physics department of Munich
university who made my stay in Germany comfortable.\newline

\end{document}